\documentclass[11pt,a4paper]{article}
\usepackage{jheppub}
\usepackage[english]{babel}
\usepackage[utf8]{inputenc}
\usepackage{enumitem}
\usepackage{amsfonts}
\usepackage{enumitem}
\setlist{nolistsep}

 \def\frac#1#2{{#1\over #2}}

 \def\s{\sqrt}

\def\be{\begin{equation}}
\def\ee{\end{equation}}
\def\ba{\begin{eqnarray}}
\def\ea{\end{eqnarray}}

 \def\f {\frac}
 \def\ti{\tilde}
 \def\ap{\alpha}

 \def\no{\nonumber \\}

 \def\la{\langle}
 \def\lb{\rangle}

\def\dpar{\vspace{11pt}}
\def\qpar{\vspace{22pt}}

\newcommand{\tr}[1][]{ \text{tr}_{#1}\; }
\begin{document}

\begin{titlepage}

\begin{flushright}
YITP-17-38\\
IPMU17-0058\\
\end{flushright}

	\centering
%	{\Large\scshape Draft from \today}
%	\vspace{1.5cm}
	
	{\parbox[c]{0.75\textwidth}{\Large\sffamily\centering \qpar Holographic Entanglement Entropy of Local Quenches in AdS$_4$/CFT$_3$: A Finite-Element Approach}\qpar\qpar}
	
	{\Large Alexander Jahn$^a$ and Tadashi Takayanagi$^{b,c}$ \dpar}
	
	\parbox[c]{0.9\textwidth}{\large\it\centering $^a$ Dahlem Center for Complex Quantum Systems, Freie Universität Berlin, 14195 Berlin, Germany \\
	$^b$ Yukawa Institute for Theoretical Physics (YITP), Kyoto University, Kyoto 606-8502, Japan\\
$^{c}$ Kavli Institute for the Physics and Mathematics of the Universe (Kavli IPMU), University of Tokyo, Kashiwa, Chiba 277-8582, Japan\qpar\qpar}
	
	{\bfseries\large Abstract \dpar}
	
	\parbox[c]{0.9\textwidth}{\large Understanding quantum entanglement in interacting higher-dimensional conformal field theories is a challenging task, as direct analytical calculations are often impossible to perform. With holographic entanglement entropy, calculations of entanglement entropy turn into a problem of finding extremal surfaces in a curved spacetime, which we tackle with a numerical finite-element approach. In this paper, we compute the entanglement entropy between two half-spaces resulting from a local quench, triggered by a local operator insertion in a CFT$_3$. We find that the growth of entanglement entropy at early time agrees with the prediction from the first law, as long as the conformal dimension $\Delta$ of the local operator is small. 
	Within the limited time region that we can probe numerically, we observe deviations from the first law and a transition to sub-linear growth at later time. In particular, the time dependence at large $\Delta$ shows qualitative differences to the simple logarithmic time dependence familiar from the CFT$_2$ case. 
	 We hope that our work will motivate further studies, both numerical and analytical, on entanglement entropy in higher dimensions.}

\end{titlepage}
\thispagestyle{empty}
\tableofcontents
\newpage
\setcounter{page}{1}
\section{Introduction}

The discovery of the AdS/CFT correspondence \cite{Ma,Gubser:1998bc_Witten:1998qj} has precipitated a number of new research directions in theoretical physics. One of the fields that has greatly benefitted from AdS/CFT involves the study of entanglement entropy. In particular, AdS/CFT provides a geometrical method to compute entanglement entropy \cite{Ryu:2006bv,HRT} (see \cite{CHM,LM} for its derivations). Although this approach had originally been developed from ideas related to the Bekenstein-Hawking formula for black hole entropy \cite{BKLS,Sr}, it has found applications within condensed matter physics and quantum information theory. Holographic entanglement entropy provides a useful method for certain strongly coupled quantum systems, so-called holographic CFTs, which are dual to classical gravity via AdS/CFT. The holographic approach has an advantage especially in dimensions larger than two\footnote{
Throughout this paper, we generally work in a relativistic setting, with dimensions refering to space-time if not otherwise noted. We also use natural units with $c=1$.}, as the analysis of entanglement entropy in interacting higher-dimensional CFTs is quite difficult.

The non-equilibrium dynamics of entanglement entropy are a field of extensive research (for a general review, see \cite{Eisert:2008ur,Eisert:2014jea}). There has already been much progress on homogeneous excitations such as global quantum quenches, which can be analytically studied in 1+1-dimensional CFTs \cite{GQ}.  Holographic studies of entanglement entropy under global quenches \cite{AAL,Ba,HaMa,Kundu:2016cgh} have been successful even in higher dimensions. However, for local excitations in CFT, our knowledge of the behavior of entanglement entropy is highly limited, especially in higher dimensions. In this paper, we  are therefore interested in holographic entanglement entropy for a specific class of locally excited states, referred to as local quenches\footnote{
Note that there is another class of local quenches
where CFTs on two semi-infinite systems are instantaneously joined together \cite{Calabrese:2007,Calabrese:2016xau}.
We will not discuss this class of local quenches in this paper.}, in 2+1-dimensional CFTs using AdS$_4/$CFT$_3$. In the CFT description, we are considering excited states $|\Psi\lb$ defined by acting with a local operator $O(x)$ on the CFT vacuum $|0\lb$ in the manner
\be
|\Psi\lb={\cal N}e^{-\ap H}O(x)|0\lb, \label{lopw}
\ee
where ${\cal N}$ is the normalization factor to unit norm. The parameter $\ap>0$ provides a UV regularization as the literally point-like localized operator has infinite energy and is singular.
We are interested in the time evolution of the entanglement entropy $S_A$ for the excited state $|\Psi(t)\lb=e^{-iHt}|\Psi\lb$ when we choose the subsystem $A$ to be the half-space. The excitation is located on the boundary between both half-spaces, thus producing additional entanglement between them. As time increases, a larger causal region is affected by the quench. Our main quantity of interest is the resulting growth of  entanglement entropy $\Delta S_A(t)=S_A(|\Psi(t)\lb)-S_A(|0\lb)$ compared to the vacuum.

Previous analyses of $\Delta S_A$ for massless scalar fields have been performed in \cite{Nozaki:2014hna,Nozaki:2014uaa,Nozaki:2015mca,Nozaki:2016mcy} and it was found that the growth $\Delta S_A(t)$ approaches a finite positive constant at late time. This is interpreted as a system of entangled particles propagating at the speed of light (see the recent discussion \cite{Nozaki:2017hby}). The same behavior has been found for rational CFTs in two dimensions \cite{Nozaki:2014uaa,He:2014mwa,Chen:2015usa,
Caputa:2015tua,Caputa:2016yzn,Numasawa:2016kmo}.
Furthermore, a recent study of 1+1-dimensional orbifold CFTs found an exotic time evolution $\Delta S_A\propto \log\log t$ for irrational CFTs \cite{Caputa:2017tju}. For other field theoretic progress on local quenches refer also to \cite{Shiba:2014uia,Caputa:2014eta,deBoer:2014sna,Guo:2015uwa,Caputa:2015waa,Rangamani:2015agy,David:2016pzn,
Sivaramakrishnan:2016qdv,Numasawa:2016emc}.

However, holographic results have so far been limited to the AdS$_3/$CFT$_2$ setup, where we can analytically compute $\Delta S_A$ \cite{Nozaki:2013wia,Caputa:2014vaa}. In this holographic description, the local excitation corresponds to a massive particle falling in AdS$_3$, whose mass $m$ is related to the conformal dimension $\Delta$ of the local operator $O(x)$ in (\ref{lopw}) via the standard relation $\Delta\simeq mR$, with $R$ being the AdS radius. The holographic results for 1+1-dimensional CFT show that $\Delta S_A\propto \log t$ under time evolution at late time \cite{Nozaki:2013wia,Caputa:2014vaa}. This time dependence has been precisely reproduced in \cite{Asplund:2014coa} using a large $c$ CFT analysis.
Such a behavior is assumed to stem from the chaotic nature of holographic CFTs, where the quasi-particle picture breaks down.

The main purpose of this paper is to conduct analogous holographic computations for AdS$_4/$CFT$_3$
(see also \cite{Rangamani:2015agy} for different perspectives on this problem).
Unfortunately, perturbative results on the AdS side are not useful for several interesting cases when we consider entanglement between two large regions. Thus, we would like to obtain an exact result for such limits and compare it with the lower-dimensional counterpart.

Our numerical approach relies on a finite-element optimization strategy which approximates extremal surfaces, required for the calculation of holographic entanglement entropy, by a discrete mesh. In contrast to previous studies using the finite-element method \cite{Fonda:2014cca}, we require a method that does not restrict the extremal surface to a timeslice (when it becomes equivalent to a minimal surface), but can find more complicated space-like solutions. This is neccessary to tackle problems without timeslice constraints, such as the holographic local quench model, where no useful Killing symmetry of the time-dependent gravity background exists.

This paper consists of three parts. In section \ref{S_2}, we will review the holographic local quench model and its solutions in $\text{AdS}_3/\text{CFT}_2$. Section \ref{S_3} will present the application of our numerical approach to the $\text{AdS}_4/\text{CFT}_3$ case and the computational results obtained with it. An interpretation of these results in terms of more general principles of entanglement entropy will be given in section \ref{S_4}. The details of our numerical method will be described in appendix \ref{APP_A}.

\newpage
\section{Entanglement Entropy of Local Quenches}
\label{S_2}
\subsection{Local quenches in AdS/CFT}

A local excitation in a CFT can be described by a holographic dual consisting of a freely falling mass in AdS Poincaré coordinates \cite{Nozaki:2013wia}, based on the construction \cite{Horowitz:1999gf}. Pure (i.e.\ ``empty'') AdS$_{d+1}$ spacetime in Poincaré coordinates $(t,z,x_1,\dots,x_{d-1})$ corresponds to
\begin{equation}
\label{EQ_POINCARE_EMPTY}
\text{d}s^2 = \frac{R^2}{z^2} \left( \text{d}z^2 - \text{d}t^2 + \sum_{i=1}^{d-1} \text{d}x_i^2 \right) \text{ .}
\end{equation}
In this notation, $R$ is the AdS radius. In such a geometry, a falling mass $m$ which is at rest at time $t=0$ at position $(z,x_1,\dots,x_{d-1})=(\alpha,0,\dots,0)$ follows a trajectory
\begin{equation}
\label{EQ_MASS_TRAJ}
z(t)=\sqrt{t^2 + \alpha^2} \text{ ,}\quad x_1(t)=\dots=x_{d-1}(t)=0 \text{ .}
\end{equation}
Such a falling mass has a conserved energy $E$, which can be most easily evaluated as the rest energy at $t=0$:
\begin{equation}
\label{EQ_QUENCH_ENERGY}
E = \frac{R}{z(t=0)} m = \frac{R}{\alpha} m \text{ .}
\end{equation}
Note that $z(t)/R$ is equivalent to an energy scale, or inverse length scale, of pure AdS spacetime.

The induced metric following from the insertion of a mass into AdS$_{d+1}$ spacetime is quite complicated. Fortunately, as found in \cite{Horowitz:1999gf}, we can express it more conveniently by switching to global coordinates $(\tau, r, \theta_1,\dots,\theta_{d-1})$ with a new time coordinate $\tau$, radius $r$ and $d-1$ angular coordinates $\theta_i$. For the pure AdS case, this corresponds to
\begin{equation}
\text{d}s^2 =  - \left(R^2+r^2\right) \text{d}\tau^2 + \frac{R^2\text{d}r^2}{R^2+r^2} + r^2 \text{d}\Omega^2 \text{ .}
\end{equation}
Here, $\text{d}\Omega$ is the standard angular differential on $S^{d-1}$, e.g.\ $\text{d}\Omega^2=\text{d}\theta_1^2 + \sin^2(\theta_1)\text{d}\theta_2^2$ for $d=3$. The new time coordinate $\tau$ is unitless. The general coordinate transformation between Poincaré and global coordinates is given by
\begin{equation}
\begin{aligned}
T \; &= \; \sqrt{R^2 + r^2} \sin{\tau}&&= \frac{R t}{z} \text{ ,} \\
W \; &= \; \sqrt{R^2 + r^2} \cos{\tau}&&= \frac{1}{2z} \left( e^\beta R^2 + e^{-\beta} \left(z^2 - t^2 + \sum_i x_i^2\right) \right) \text{ ,} \\
Z \; &= \;  r \Omega_d & &= \frac{1}{2z} \left( -e^\beta R^2 + e^{-\beta} \left(z^2 - t^2 + \sum_i x_i^2\right) \right) \text{ ,}\\
X_i \; &= \; r \Omega_i & &= \frac{R x_i}{z} \text{ ,}
\label{EQ_GLOBALCOORDS_3}
\end{aligned}
\end{equation}
where we expressed the transformation in terms of the usual AdS embedding coordinates $(T,W,Z,X_1,\dots,X_{d-1})$, in a $\mathbb{R}^{d,2}$ manifold. The expressions $\Omega_1,\dots,\Omega_d$ are the Cartesian components of the unit vector in terms of the angular coordinates, e.g.\ $\Omega_1=\cos\theta_1 \sin\theta_2$ for $d=2$. Note that the AdS boundary, located at $z=0$ in Poincaré coordinates, appears at $r=\infty$ in global coordinates.

The real parameter $\beta$, corresponding to an additional boost transformation, does not appear in the invariant $\text{d}s^2$ in the pure AdS case. This changes when we consider the additional mass $m$: If we choose $e^\beta = \frac{\alpha}{R}$, the trajectory \eqref{EQ_MASS_TRAJ} corresponds to a static point at $r=0$ in global coordinates. Thus, the induced metric in global coordinates is that of a static AdS black hole, independent of global time $\tau$. Explicitly, the full solution \cite{Witten:1998zw} is given by
\begin{equation}
\label{EQ_GLOBAL_FULL}
\text{d}s^2 = - (R^2 + r^2 - \frac{M}{r^{d-2}}) \text{d}\tau^2 + \frac{R^2}{R^2 + r^2 - \frac{M}{r^{d-2}}} \text{d}r^2 + r^2 \text{d}\Omega_{d-1}^2 \text{ ,}
\end{equation}
where the parameter $M$ is related to the mass $m$ via
\begin{equation}
\label{EQ_ADS_MASS}
M = \frac{8 \Gamma(\frac{d}{2}) G_N R^2}{(d-1) \pi^{d/2-1}} m \text{ ,}
\end{equation}
where $G_N$ is the Newton constant (gravitational constant).
We can use the coordinate transformation \eqref{EQ_GLOBALCOORDS_3} to map the solution \eqref{EQ_GLOBAL_FULL} back to Poincaré coordinates $(t,z,x_1,\dots,x_{d-1})$. However, the resulting expression for the invariant $\text{d}s^2$ is rather involved and is therefore omitted here.

\begin{figure}[t]
\hspace{5.5pt}
\includegraphics[scale=0.3]{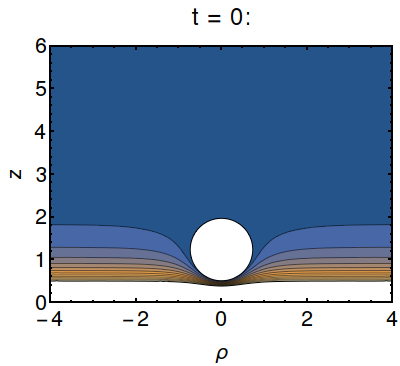}
\hspace{5.2pt}
\includegraphics[scale=0.3]{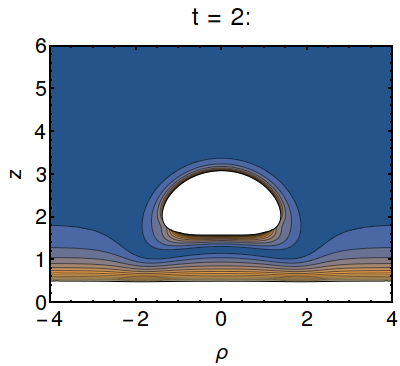}
\hspace{5.7pt}
\includegraphics[scale=0.299]{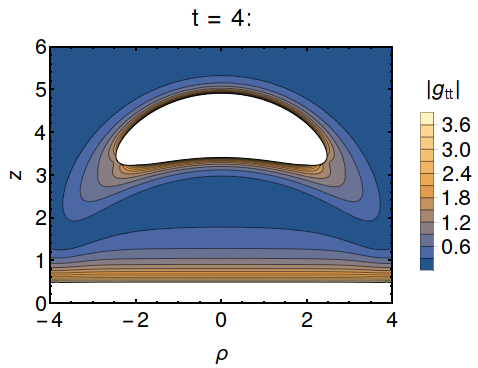} \\
\includegraphics[scale=0.323]{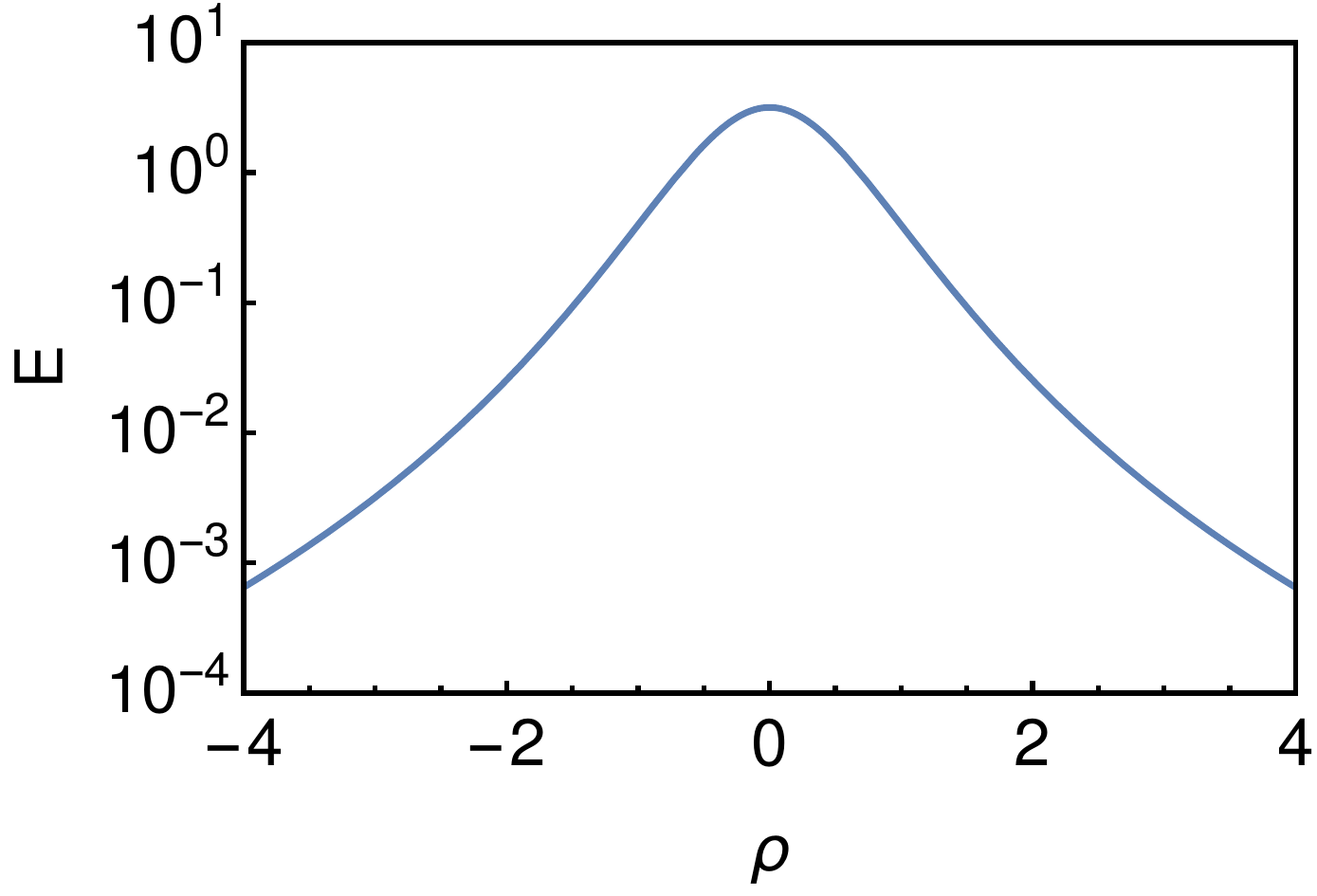}
\includegraphics[scale=0.323]{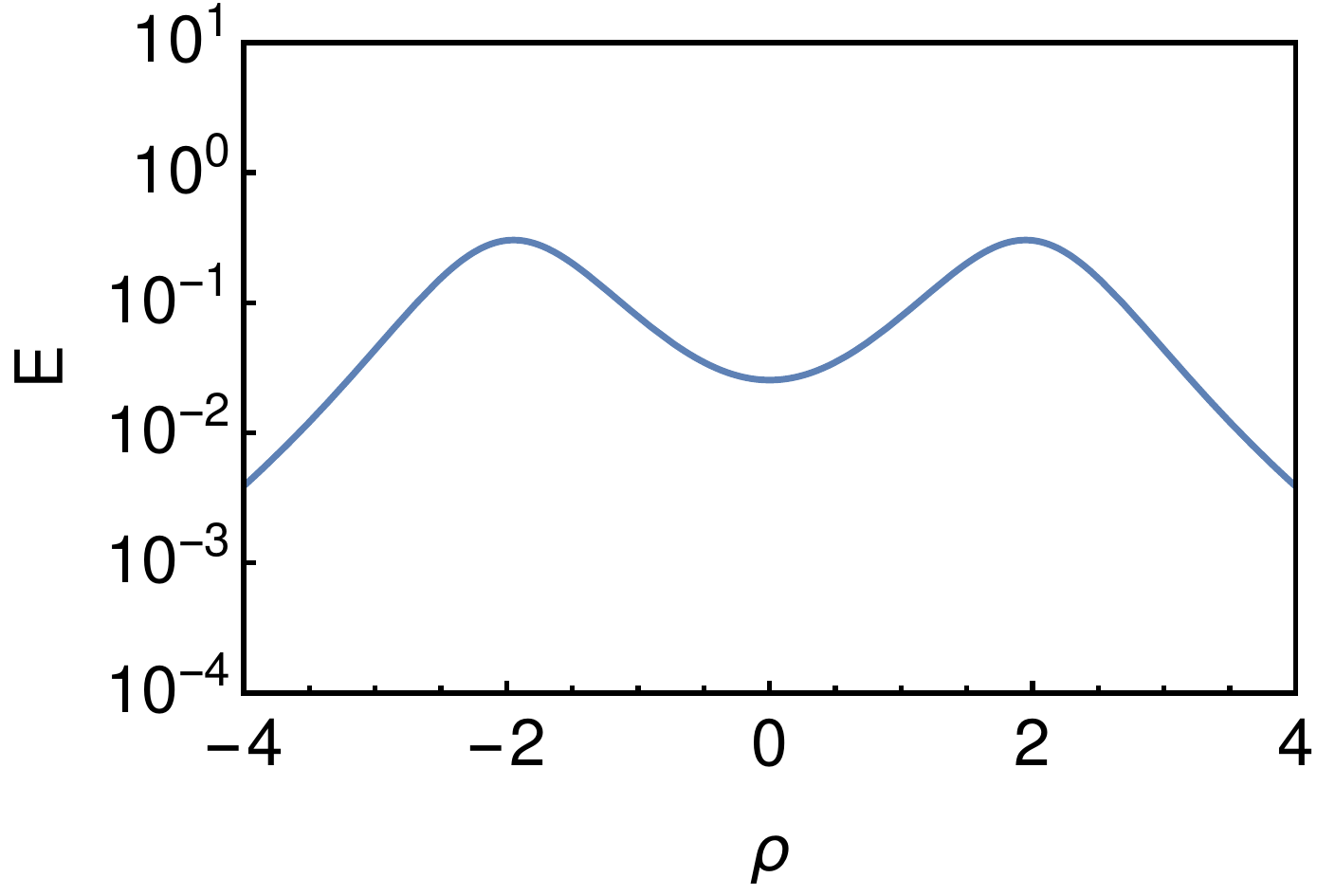}
\includegraphics[scale=0.323]{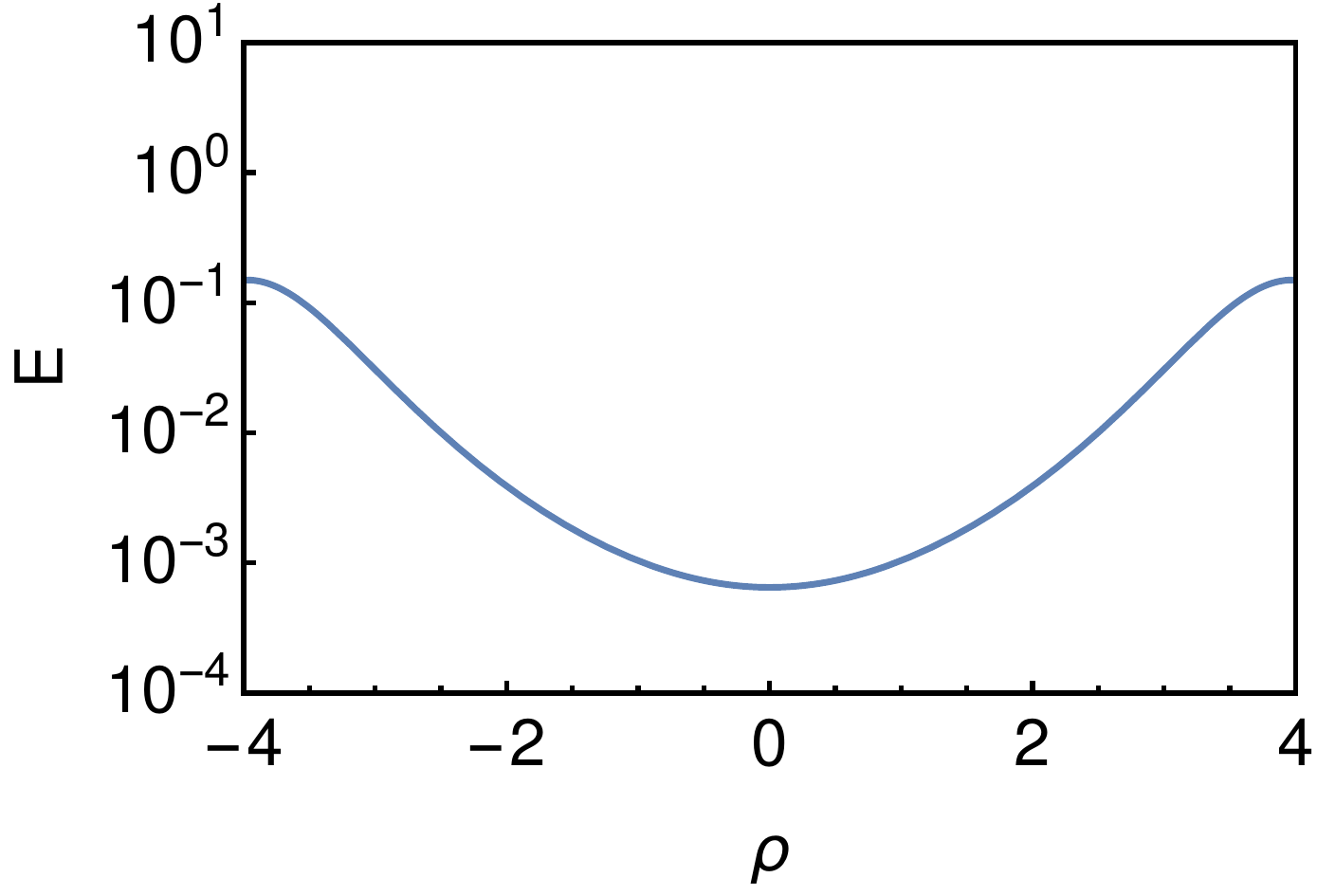}
\caption{Absolute value of the metric component $g_{t t}$ in Poincaré coordinates (top) and corresponding boundary energy density $E=T_{t t}$ (bottom) for time $t=0,2,4$ in the AdS$_4$/CFT$_3$ local quench model. Horizontal axis shows radius $\rho=\sqrt{x_1^2 + x_2^2}$, boundary is at $z=0$. Units correspond to $M=R=\alpha=1$.}
\label{FIG_METRIC_CONTOUR}
\end{figure}

The falling mass in Poincaré AdS$_{d+1}$ corresponds to a local excitation in the CFT$_d$ on the boundary. This is visualized in figure \ref{FIG_METRIC_CONTOUR} for the case $d=3$: We show the component $g_{t t}$ of the induced metric in Poincaré coordinates as well as the corresponding boundary energy density, which is given by the $T_{t t}$ component of the stress energy tensor at $z\to 0$ (for details on the procedure, see \cite{Nozaki:2013wia}). Introducing the radial coordinate $\rho=\sqrt{x_1^2 + x_2^2}$, the latter is given by
\begin{equation}
\label{EQ_CFT3_T_TT}
T_{t t}(\rho,t) = \frac{M \alpha^3}{\pi G_N R}\frac{(\alpha^2+t^2+\rho^2)^2 + 2 t^2 \rho^2}{\left((\alpha^2 + t^2-\rho^2)^2 + 4 \alpha^2 \rho^2\right)^{5/2}} \text{ .}
\end{equation}
As we can see, the falling mass in the AdS$_4$ bulk creates complicated deviations from the pure AdS metric, which correspond to a radially symmetric excitation on the boundary starting at $t=0$. The peak of this shockwave has a width $\sim \alpha$ and approaches the speed of light at $t \gg \alpha$. This falling particle description can be applied to local quenches in dimensions $d$ larger than three, as well \cite{Nozaki:2013wia,Nozaki:2014uaa}.

The falling particle corresponds to a localized state in the AdS bulk, so we expect entanglement between any two regions on the peak of the shockwave. Thus, it is compelling to take a closer look at the entanglement properties of the local quench system.

\subsection{Half-space entanglement entropy}

The entanglement entropy $S_A$ of a subsystem $A$ of a CFT quantifies the entanglement between $A$ and its conjugate $B=A^C$ of the CFT space. By using the holographic entanglement entropy formula \cite{Ryu:2006bv,HRT}, it can be calculated as
\begin{equation}
\label{EQ_RYUTAKAYANAGI}
S_A = \frac{|\gamma_A|}{4 G_N} \text{ ,}
\end{equation}
where $\gamma_A$ is an extremal hypersurface along the boundary of $A$ (i.e.\ $\partial \gamma_A = \partial A$) extending into the AdS bulk, and $|\gamma_A|$ is its area. For the AdS$_{d+1}$/CFT$_d$ setup, $\gamma_A$ has $d-1$ dimensions.

For the case of a local quench, an interesting choice of $A$ is that of a half-space in Poincaré coordinates, given by
\begin{equation}
\label{EQ_HALFSPACE}
A = \left\lbrace (t=t_0,\; z=z_0,\; x_1,\dots,x_{d-1}) \quad | \quad x_1 \in \mathbb{R}_{>0} \;\wedge\; x_2,\dots,x_{d-1} \in \mathbb{R} \right\rbrace \text{ ,}
\end{equation}
where we consider the system at constant time $t_0$ and UV cutoff $z_0$. For $d=2$, the setup resembles the joining of two half-line subsystems at $x_1=0$, with $S_A$ characterizing the entanglement between both subsystems after they have been connected. For $d>2$, it corresponds to the entanglement between two half-spaces after a point-like excitation on the boundary between them.

In the $d=2$ case, $S_A$ can be analytically calculated using \eqref{EQ_RYUTAKAYANAGI}. Consider a compact region $A^\prime$ bounded by the points $(t_0,z_0,x_1)$ and $(t_0,z_0,x_2)$ in Poincaré coordinates $(t,z,x)$. Clearly, $A^\prime \to A$ as $(x_1,x_2) \to (0, \infty)$. In global coordinates, the boundary points correspond to:
\begin{equation}
\begin{aligned}
\label{EQ_ADS3_COORD_TRANS}
\tan{\tau_i} &= \frac{2 t_0 / \alpha}{(x_i^2 - t_0^2)/\alpha^2 + 1} \text{ ,}\\
\tan{\theta_i} &= \frac{2 x_i / \alpha}{(x_i^2 - t_0^2)/\alpha^2 + 1}  \text{ ,}\\
r_i &= \frac{R}{2 z_0/\alpha} \sqrt{ 4 x_i^2/\alpha^2 + \left((x_i^2 - t_0^2 )/\alpha^2 - 1\right)^2} \text{ ,}
\end{aligned}
\end{equation}
with $i=1,2$. Note that the constant $\alpha$ acts as a scaling factor in Poincaré coordinates, while $R$ corresponds to a radial rescaling in global coordinates. Also, $r_1$ and $r_2$ diverge as the points approach the AdS boundary at $z_0 = 0$.

The entanglement entropy for this subsystem $A$ is given by \cite{Nozaki:2013wia}
\begin{equation}
\label{EQ_ADS3_DELTA_SA_GENERAL}
S_A = \frac{R}{4 G_N} \left[ \log\left( \frac{2}{s} r_1 r_2 \right) + \log \left( \cos \left(\frac{\sqrt{s}}{R} |\tau_2 - \tau_1| \right)- \cos\left(\frac{\sqrt{s}}{R} |\theta_2 - \theta_1| \right)  \right) \right] \text{ ,}
\end{equation}
with $s=R^2-M$. For $s<0$, the expression can be analytically continued so that $S_A$ is real. We have omitted terms of order $O\left(r_1^{-2}, r_2^{-2}\right)$ that vanish on the AdS boundary at $r_1,r_2 \to \infty$. In this limit, the first term is logarithmically divergent. Therefore, we rely on the finite quantity $\Delta S_A = S_A - \left. S_A \right|_{M=0}$ to describe the system, giving the entanglement entropy excited by the local quench.

The half-line subsystem we are interested in corresponds to setting $x_1=0$ and considering the limit $x_2 \to \infty$. First, consider the $t=0$ case, where the entanglement entropy becomes
\begin{equation}
\label{EQ_ADS3_DELTA_SA_T0}
\Delta S_A(t=0) = \frac{R}{2 G_N} \log \left( \frac{\sin\left( \frac{\pi}{2} \sqrt{1 - M/R^2} \right)}{\sqrt{1 - M/R^2}} \right) \text{ .}
\end{equation}

As we are mainly interested in the region $t \gg \alpha$ where the two peaks of the CFT excitation are clearly separated, we can also expand \ref{EQ_ADS3_DELTA_SA_GENERAL} in $\frac{\alpha}{t}$, yielding
\begin{align}
\label{EQ_ADS3_DELTA_SA}
\Delta S_A(t) &= \frac{R}{4 G_N} \log \left( \frac{\sin\left(\pi\sqrt{1 - M/R^2} \right)}{\sqrt{1- M/R^2}} \frac{t}{\alpha} - \cos\left(\pi\sqrt{1 - \frac{M}{R^2}} \right) + O\left(\frac{\alpha}{t}\right) \right) \nonumber \\
&\simeq \frac{R}{4 G_N} \left[ \log \left( \frac{\sin\left(\pi\sqrt{1 - M/R^2} \right)}{\sqrt{1- M/R^2}} \right) + \log \frac{t}{\alpha}\right] \text{ .}
\end{align}
To produce the second line, we assumed $\sqrt{1 - M/R^2} \cot\left(\pi\sqrt{1 - M/R^2} \right) \ll \frac{t}{\alpha}$. As $M=8 G_N R^2 m$ is a nonzero parameter, this simplification is valid at large time $t$. Note that the central charge is given by the standard formula $c=\frac{3R}{2G_N}$
\cite{BrHe} and the conformal dimension of the local operator is expressed as $\Delta=\frac{M}{8G_NR}$.
This leads to the relation $M/R^2=24\Delta/c$.

Thus, a local quench produces an asymptotically logarithmic time dependence in holographic 2-dimensional CFTs \cite{Nozaki:2013wia,Caputa:2014vaa}. Furthermore, the coefficient of logarithmic growth $\frac{R}{4G_N}=\frac{c}{6}$ does not depend on $M$ (or equally the conformal dimension $\Delta$), that is, the size of the excitation. This logarithmic behavior was reproduced precisely in the 2-dimensional CFT analysis of \cite{Asplund:2014coa}, using the large central charge method. It is curious to note that a similar logarithmic behavior has also been found in CFT calculations for a different class of local quenches created by joining two half-lines \cite{Calabrese:2007}, though the coefficient is different by a factor of two. Note also that the dependence on time $t$ in the form $\frac{t}{\alpha}$ follows directly from the invariance of the system under $(t,z,x) \to (a\, t,a\, z, a\, x)$ and $\alpha \to a\, \alpha$ for any $a>0$.

It might be intriguing to compare the above results for holographic
CFTs with those for integrable CFTs, where field theoretic computations can be done analytically.
For 2-dimensional rational CFTs, including free CFTs, the growth of entanglement $\Delta S_A$ is finite and becomes a step function in the limit $\ap\to 0$, which is simply explained by the behaviour of entangled particles propagating at the speed of light \cite{Nozaki:2014hna,Nozaki:2014uaa,He:2014mwa,Chen:2015usa,
Caputa:2015tua,Caputa:2016yzn,Numasawa:2016kmo}. 
Recently, another class of integrable 2-dimensional CFTs, namely orbifold CFTs, has been studied \cite{Caputa:2017tju}. For irrational CFTs, an exotic time evolution $\Delta S_A\propto \log\log t$ has been found.
Refer to \cite{Shiba:2014uia,Caputa:2014eta,deBoer:2014sna,Guo:2015uwa,Caputa:2015waa,David:2016pzn,Numasawa:2016emc} for further results for local quenches in two dimensions.

In 3-dimensional CFTs, on the other hand, the only available results for local quenches created by the local operator insertions are those for the free field CFTs \cite{Nozaki:2014hna,Nozaki:2014uaa,Nozaki:2015mca,Nozaki:2016mcy}. In free field CFTs, we can again
understand the evolution of entanglement entropy based on the picture of entangled particles propagating at the speed of light \cite{Nozaki:2014hna,Nozaki:2014uaa,Nozaki:2017hby}.
Therefore, it is desirable to explore many other examples of interacting CFTs in higher dimensions. Motivated by this, we focus on holographic CFTs in three dimensions
(CFT$_3$) in this paper. It is natural to ask whether our holographic 3-dimensional case exhibits the same logarithmic growth $\Delta S_A\propto \log t$ as the 2-dimensional one (\ref{EQ_ADS3_DELTA_SA}).

\subsection{Extension to AdS$_4$/CFT$_3$}

We will now consider the $d=3$ case of a holographic local quench. Again, we map the Poincaré coordinate description of a falling mass to global coordinates, where the induced metric \eqref{EQ_GLOBAL_FULL} takes the form:
\begin{equation}
\text{d}s^2 = - (R^2 + r^2 - \frac{M}{r}) \text{d}\tau^2 + \frac{R^2}{R^2 + r^2 - \frac{M}{r}} \text{d}r^2 + r^2 \text{d}\theta^2 + r^2 \sin^2{\theta} \; \text{d}\phi^2 \text{ .}
\end{equation}
According to \eqref{EQ_ADS_MASS}, the mass parameter $M$ is now related to the real mass $m$ via $M=2 G_N R^2 m$. A point in global coordinates $(\tau, r, \phi, \theta)$ is translated to a point in Poincaré coordinates $(t,z,x,y)$ via
\begin{equation}
\begin{aligned}
\label{EQ_ADS4_COORD_TRANS}
\tan \tau &= \frac{2 t / \alpha}{(x^2 + y^2 + z^2 - t^2)/\alpha^2 + 1} \text{ ,}  \\
r \cos\phi \sin\theta &= \frac{R x}{z} \text{ ,} \\
r \sin\phi \sin\theta &= \frac{R y}{z} \text{ ,} \\
r \cos\theta &= \frac{R}{2 z / \alpha} \left( (x^2 + y^2 + z^2 - t^2)/\alpha^2 - 1 \right)  \text{ .}
\end{aligned}
\end{equation}

The half-space subsystem \eqref{EQ_HALFSPACE} we are particularly interested in corresponds to a boundary $\partial A$ that is given by a line at $y=0$ in Poincaré coordinates. In the pure AdS case, i.e.\ for $M=0$, the extremal surface $\gamma_A$ is simply the $xz$ plane at constant time:
\begin{equation}
\gamma_A^{(0)} = \left\lbrace (t=t_0,\; z,\; x,\; y=0) \quad | \quad x,z \in \mathbb{R} \;\wedge\; z>z_0 \right\rbrace \text{ .}
\end{equation}
Even for a finite $z_0$ the surface area $|\gamma_A^{(0)}|$ is infinite. For any nonzero $M$, however, the corresponding extremal surface $\gamma_A$ will only significantly differ from $\gamma_A^{(0)}$ in some local region around the mass. Thus, the growth of entanglement entropy
\begin{equation}
\Delta S_A = \frac{|\gamma_A|-|\gamma_A^{(0)}|}{4 G_N}
\end{equation}
is well-defined and finite.

Assuming only small deformations from $\gamma_A^{(0)}$ to $\gamma_A$, we can attempt to calculate $\Delta S_A$ perturbatively. The induced metric $G_{\mu \nu}$ resulting from projecting the full metric for $M>0$ on the plane $\gamma_A^{(0)}$ can be expanded in orders of $M$. The area of $\gamma_A$ can then be approximated by the expansion
\begin{align}
\label{EQ_A_PERT}
|\gamma_A| &\simeq \int \text{d}^2\xi \sqrt{\det (G^{(0)} + G^{(1)} + O(M^2))}  \nonumber \\
&=  \underbrace{\int \text{d}^2\xi \sqrt{\det G^{(0)}}}_{|\gamma_A^{(0)}|} +   \underbrace{\frac{1}{2} \int \text{d}^2\xi \sqrt{\det G^{(0)}} \text{Tr} [ G^{(1)} (G^{(0)})^{-1} ]}_{|\gamma_A^{(1)}|} \; + \; O(M^2) \text{ ,}
\end{align}
where $\xi$ denotes coordinates parameterizing $\gamma_A^{(0)}$. The first term of this expansion becomes irrelevant, as we can now write
\begin{equation}
\Delta S_A \simeq \frac{|\gamma_A^{(1)}|}{4 G_N} \text{ .}
\end{equation}
Evaluating this expression for the full $M>0$ metric on the $xz$ plane and  considering the limit $t \ll \alpha$, we find that
\begin{equation}
|\gamma_A^{(1)}| = \frac{1}{2} \int_0^\infty \text{d}z \int \text{d}^{2}x \frac{R}{z} \operatorname{Tr} G^{(1)} \simeq \frac{4 M t}{R \alpha} \text{ .}  \label{lqxx}
\end{equation}

This perturbative approach suggests a linear growth of the entanglement entropy with time $t$ for small perturbations in the minimal surface. However, this result rests on the assumption that at small $M$, the growth of entanglement entropy is dominated less by changes in the shape of $\gamma_A$ than by changes in the background metric. Indeed this approximation corresponds to
the first law relation of entanglement entropy, which can be applied only for small excitation, as we will explain in section \ref{S_4}.

%In fact, this perturbative result is plainly unreliable in the AdS$_3$ case, where a similar calculation also predict $\Delta %S_A \sim t$, directly contradicting \eqref{EQ_ADS3_DELTA_SA}.

Thus, we desire a non-perturbative approach to calculating the extremal surface area $|\gamma_A|$. However, an analytic computation holds considerable challenges. Even for the simple case of a disk-shaped subsystem $A$ around $(x,y)=(0,0)$, where \eqref{EQ_ADS4_COORD_TRANS} allows us to constrain the solution to constant $\tau$, computing $\gamma_A$ requires the evaluation of integrals that do not have an analytic expression. Solutions for the half-space subsystem are even more involved, motivatating the use of a numerical approach. Note that this situation differs from the AdS$_3$ case, where we can compute the holographic entanglement entropy analytically. Similarly, analytical results can be obtained from field theoretic computations only in two dimensions. Therefore a non-perturbative AdS$_4$ analysis using numerical tools allows us to make predictions which are impossible in any field theoretic analysis currently available.

\section{Numerical Studies in AdS$_4$/CFT$_3$}
\label{S_3}
\subsection{Numerical surface extremization}

Finding an optimal surface is a common numerical problem, often tackled using a finite element discretization. This discretization, usually a triangulation or quadrilateralization, approximates a continuous surface by a finite number of parameters that can be varied until a solution is found that optimizes a function of these parameters. By recursively refining the discretization, i.e.\ enlarging the parameter space of the optimization, a series of discretized approximations converging to the continuous solution is produced.

When searching for an extremal surface, the optimization function is the area of the surface itself. Such optimization methods are frequently used, and have even been applied to minimal surfaces in pure AdS$_4$ spacetime \cite{Fonda:2014cca}, which are related to ground state entanglement entropies. However, in our case of time-dependent backgrounds, we need to find extremal surfaces, i.e.\ space-like surfaces minimal with regard to space-like variations and maximal with regard to time-like ones. This introduces considerable complications compared to simple minimization problems. In particular, it imposes highly nonlinear constraints on the parameter space of discretized solutions, as each discretization element has to remain purely space-like.

The details of our implementation are shown in appendix \ref{APP_A}. In principle, it can be used to calculate extremal surfaces for a wide range of boundary conditions and 4-dimensional metrics.

\subsection{Computation of half-space entanglement entropy}

We calculate the growth of entanglement entropy $\Delta S_A = (|\gamma_A|-|\gamma_A^{(0)}|)/4 G_N$ produced by a falling mass in the bulk AdS$_4$ spacetime. As described in the previous section, the extremal surface $\gamma_A$ is computed for a given $M$, while $\gamma_A^{(0)}$ is the flat extremal surface for $M=0$. The numerical computation is performed in global coordinates $(\tau, r, \phi, \theta)$\footnote{
As explained in \ref{ss_timeslice_constraints}, we actually use a modified global time coordinate $\tau^\prime$ which is constant on the boundary.} with a radial coordinate $r$ and a static spherical horizon at $r=r_\text{hor}(M)$. $\gamma_A^{(0)}$ is equal to the $XZ$ plane, where $X=r \sin\theta \cos\phi$ and $Z=r \cos\theta$. Because the areas of both surfaces are divergent, a cutoff parameter is required. Notice that $\gamma_A$ turns into $\gamma_A^{(0)}$ at large $r$, far away from the horizon. Thus, we should choose our cutoff along $\gamma_A^{(0)}$, i.e.\ on the $XZ$ plane.

\begin{figure}[tb!]
\centering
\includegraphics[width=0.72\textwidth]{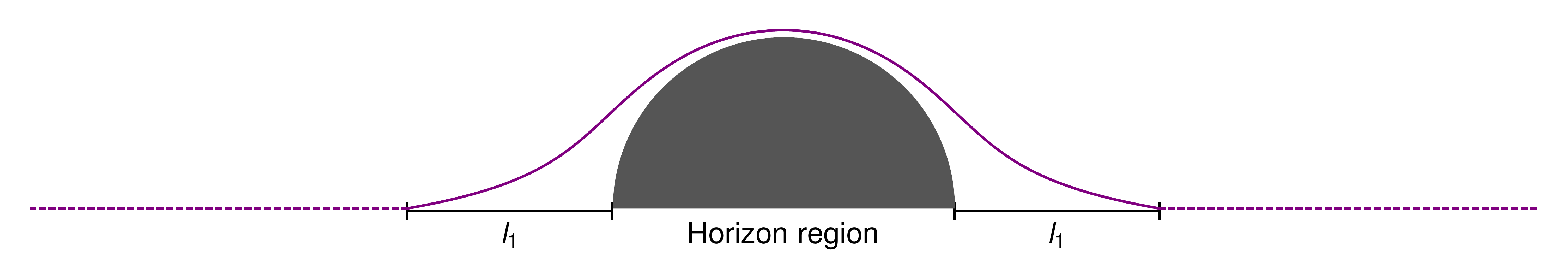}
\includegraphics[width=0.72\textwidth]{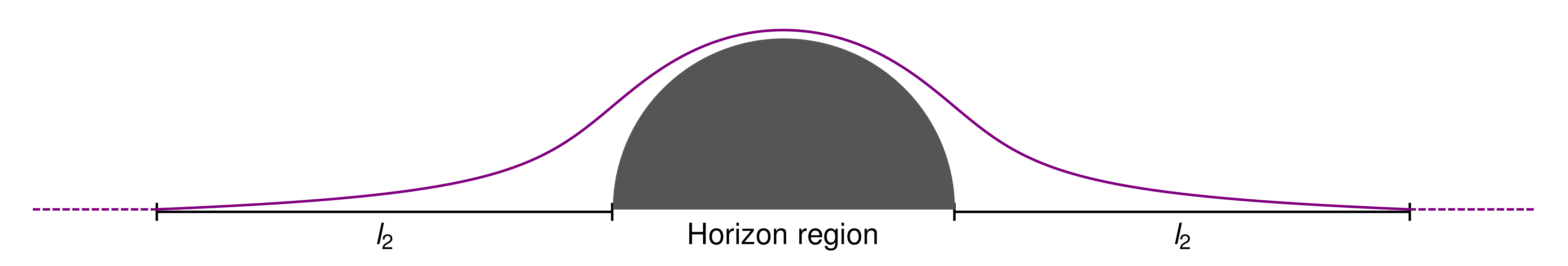}
\caption{Schematic boundary cutoff for an extremal surface arond a horizon region. Curve represents full numerical solution, dashed line corresponds to static integration region (extends to infinity in both directions). The geodesic lengths $l_1$ and $l_2$  between the horizon and the boundary of the full solution serve as an effective cutoff distance.}
\label{FIG_BOUNDARY_CUTOFF_SCHEMATIC}
\end{figure}

\begin{figure}[htb!]
\centering
\begin{minipage}{0.3\textwidth}
\centering
$\mathsf{Global, l_1=1.5:}$\\
\framebox{\includegraphics[width=0.85\textwidth]{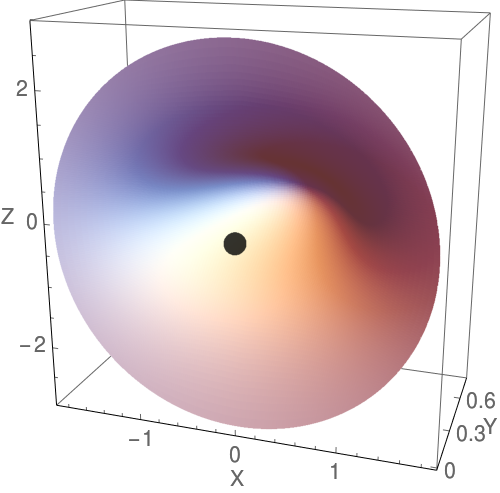}}\\
\vspace{3pt}
$\mathsf{Poincar\acute{e}, l_1=1.5:}$\\
\framebox{\includegraphics[width=0.85\textwidth]{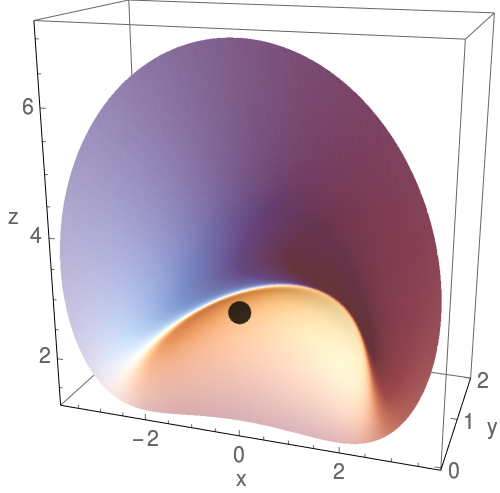}}
\end{minipage}
\begin{minipage}{0.3\textwidth}
\centering
$\mathsf{Global, l_2=2.0:}$\\
\framebox{\includegraphics[width=0.85\textwidth]{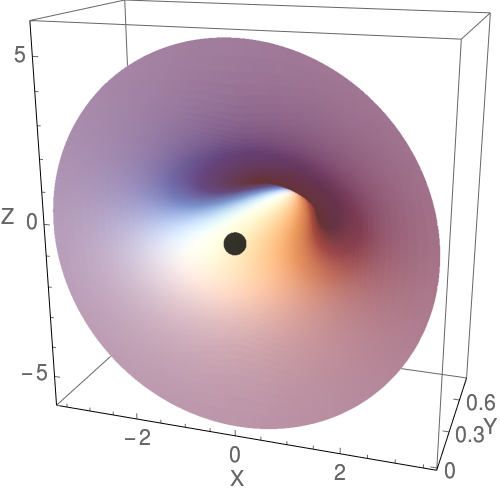}}\\
\vspace{3pt}
$\mathsf{Poincar\acute{e}, l_2=2.0:}$\\
\framebox{\includegraphics[width=0.85\textwidth]{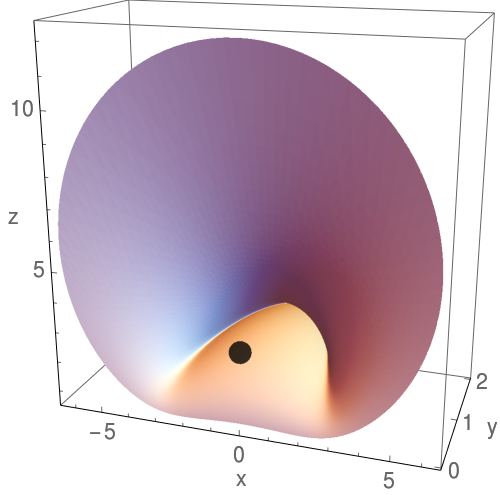}}
\end{minipage}
\caption{Example of a cutoff at proper distances $l_1=1.5$ (left) and $l_2=2.0$ (right) from the horizon, in global (top) and Poincaré coordinates (bottom). Global coordinate labels are $(X,Y,Z)=(r \sin\theta \cos\phi,r \sin\theta \sin\phi,r \cos\theta)$. The center of the coordinate horizon is shown as a black dot. Surfaces were computed for Poincaré time $t=3$ and mass parameter $M=1$. The time coordinates are omitted. Units in $R=\alpha=1$.}
\label{FIG_BOUNDARY_CUTOFF}
\end{figure}

It is natural to introduce a geodesic distance $l$ from the horizon as a cutoff parameter.  At $t=0$, the metric is completely isotropic, so the cutoff is simply a circular region at constant $r$ on the $XZ$ plane. The shortest proper distance betwen the cutoff region and the horizon is then given by $l$. Outside the cutoff, we assume $\gamma_A$ lies on the $XZ$ plane. The contribution to $|\gamma_A|-|\gamma_A^{(0)}|$ in this region can be computed using simple numerical integration. By extrapolating the convergence of $\Delta S_A$ with $l$, we can calculate the $l \to \infty$ limit. Our cutoff procedure is visualized in figure \ref{FIG_BOUNDARY_CUTOFF_SCHEMATIC} for two cutoff distances $l_1$ and $l_2$.

For $t>0$, the metric becomes increasingly anisotropic. We therefore replace the circular cutoff by an elliptic one, given by
\begin{equation}
a \,X^2 + b\,Z^2 = r^2 (a \sin^2\theta \cos^2\phi + b \cos^2\theta) = \text{const ;} \quad Y=r \sin\theta \sin\phi=0 \text{ ,}
\end{equation}
with $a$ and $b$ chosen so that the distance between the horizon and the cutoff is equal to $l$ along both the $X$ and $Z$ axes (corresponding to $(\theta,\phi)=(\pi/2,0)$ and $(0,0)$). Due to the symmetries of our metric, geodesics are still straight lines along these axes. An example for the elliptic cutoff at two different $l$ is shown in figure \ref{FIG_BOUNDARY_CUTOFF}, using both global and Poincaré coordinates.

In our definition of Poincaré and global coordinates, we introduced the AdS radius $R$ and the Poincaré distance $\alpha$ determining the starting point of the falling mass $m$. For our numerical purposes, we can set $R=\alpha=1$. As mentioned earlier, $\alpha$ acts as a scaling factor in Poincaré coordinates, thus setting $\alpha=1$ is equivalent to considering time dependence with respect to $\tilde{t}=\frac{t}{\alpha}$. For the global coordinates in which the computations are performed, $R=1$ corresponds to rescaling the metric to $\text{d}\tilde{s}^2=\frac{\text{d}s^2}{R^2}$ and using a unitless mass parameter $\tilde{M} = \frac{M}{R^3} = \frac{2 G_N m}{R}$. Thus we can write the growth of entanglement entropy in AdS$_4$/CFT$_3$ in the form
\begin{equation}
\Delta S_A = \frac{R^2}{4 G_N}\cdot \Delta\tilde{A} \left(\tilde{t}, \tilde{M}\right) \text{ ,} \label{dsjan}
\end{equation}
where $\Delta\tilde{A}(\tilde{t}, \tilde{M})$ is a unitless function of unitless parameters, to be determined numerically. In the following sections, we usually omit the rescaled units and express the results directly in terms of $\Delta S_A$. Note that in terms of CFT quantities, we can write
\be
\ti{M}=\frac{\Delta}{c_{3d}} \text{ .}
\ee
In this equation, $\Delta$ is the conformal dimension of local operator $O(x)$ (\ref{lopw}) and
$c_{3}\equiv\frac{R^2}{2G_N}$ is a conventional measure of the degrees of freedom of the 3-dimensional holographic CFT, which is a generalization of the central charge in 2-dimensional CFTs.
Thus in the CFT language we can write (\ref{dsjan}) as
$\Delta S_A=\frac{c_{3}}{2}\cdot \Delta \ti{A}\left(\ti{t},\f{\Delta}{c_{3}}\right)$.

\subsection{Numerical results}

At time $t=0$, when the mass is at rest in Poincaré coordinates, we can ignore variations in $t$ and need only to find a spatially minimal surface on a timeslice. This is because time reversal symmetry of the trajectory of the mass leads to a symmetry in the minimal surface in $t$. In addition, setting $t=0$ leads to a metric that is isotropic in global coordinates.

\begin{figure}[htb!]
\centering
\includegraphics[width=0.5\textwidth]{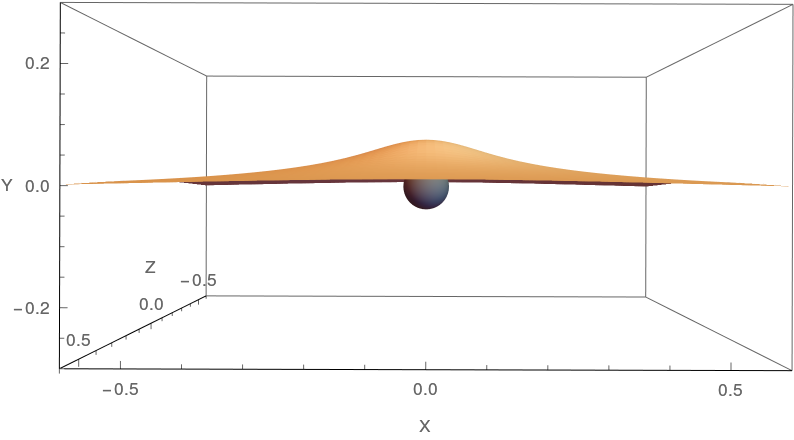}
\includegraphics[width=0.5\textwidth]{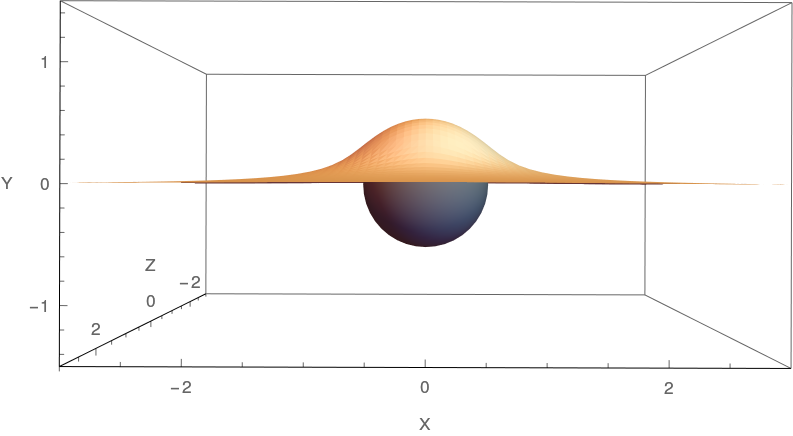}
\includegraphics[width=0.5\textwidth]{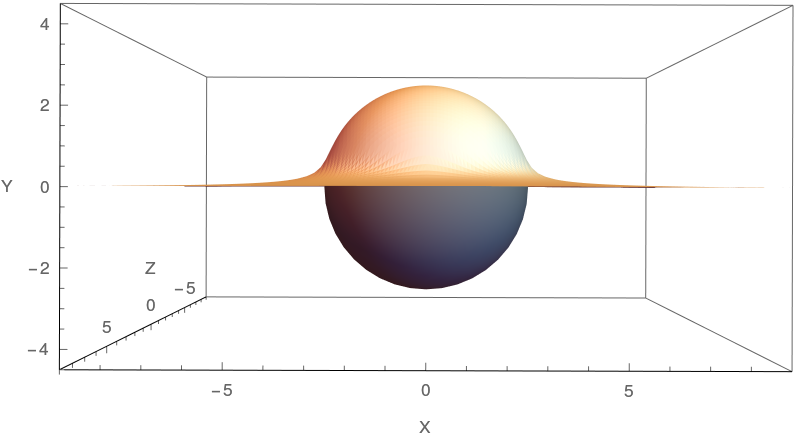}
\caption{Numerically minimized surfaces in global coordinates at quench time $t=0$ for mass parameters $M=0.05$ (top), $M=1$ (middle), and $M=40$ (bottom). The black spheres show the respective coordinate horizons of the metric. Coordinate labels are $(X,Y,Z)=(r \sin\theta \cos\phi,r \sin\theta \sin\phi,r \cos\theta)$. Units in $R=\alpha=1$.}
\label{FIG_T0_SURFACES}
\end{figure}

A series of computed discretized minimal surfaces at $t=0$ for different values of $M$ is shown in figure \ref{FIG_T0_SURFACES}. At large $M$, the minimal surface ``wraps'' closely around the coordinate horizon of our metric \eqref{EQ_GLOBAL_FULL}. As this metric is only an approximation to a horizon-less metric (like that of a star) corresponding to a proper pure state, this means that results are not completely physical at very large $M$.

\begin{figure}[tb!]
\centering
\includegraphics[width=0.6\textwidth]{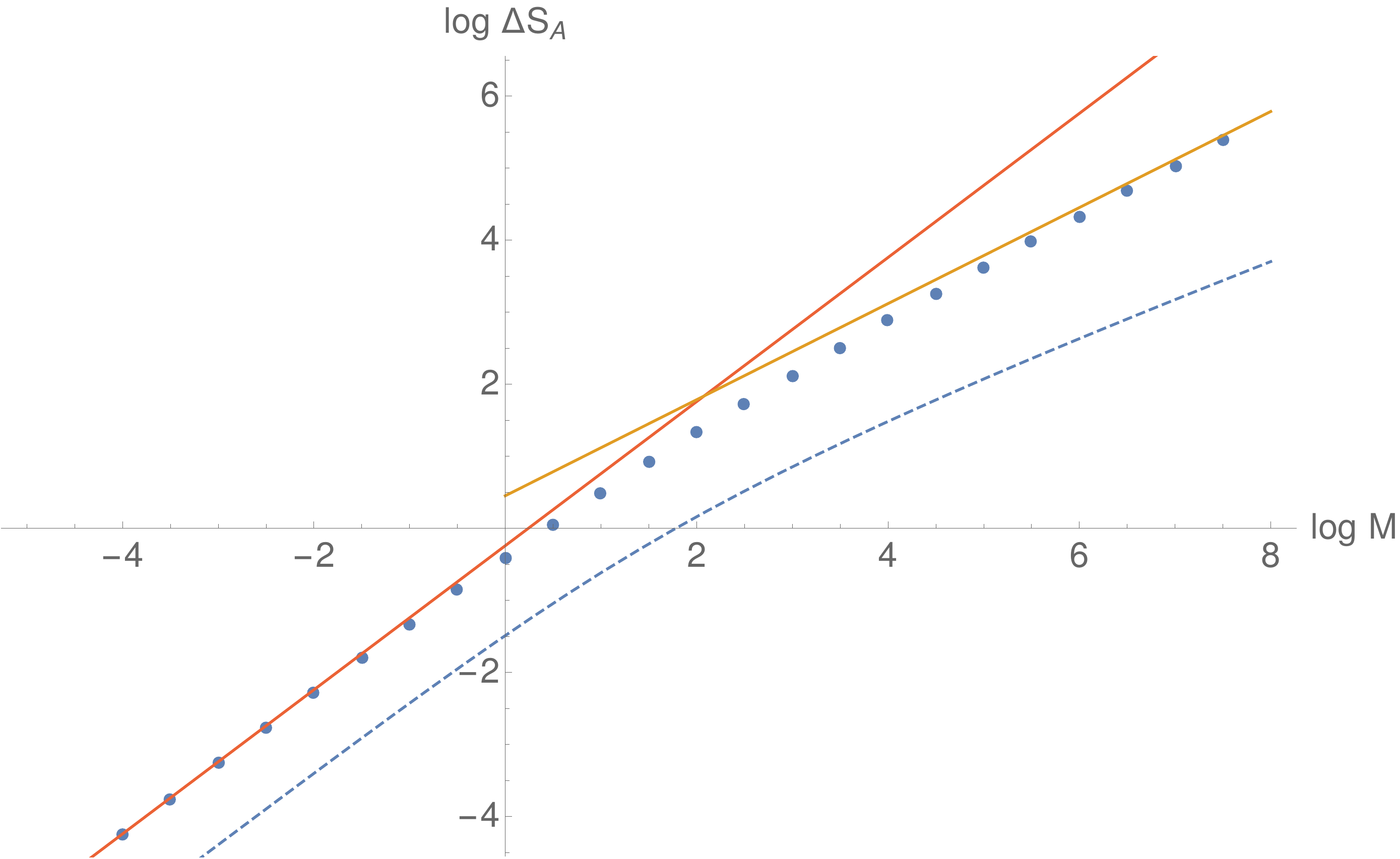}
\caption{Half-space entanglement entropy $\Delta S_A$ of a local quench at time $t=0$. Data points show numerical computations in AdS$_4$, dashed line is the analytical AdS$_3$ result. The solid lines correspond to functions $\Delta S_A(M) = \frac{\pi}{4} M$ (red) and $\frac{\pi}{2} M^{2/3}$ (orange). Units in $G_N = R = 1$.}
\label{FIG_EE_T0}
\end{figure}

The rotational invariance and time-slice constraint allow us to simplify the algorithm from appendix \ref{APP_A} considerably, as we only have to find the shape of a 1-dimensional profile curve in 2-dimensional space. Note that at $t\neq 0$, both simplifications break down and we have to use the full algorithm.

The quantitative results for $\Delta S_A$ at $t=0$ are shown in figure \ref{FIG_EE_T0}. For small $M$, the data points closely follow a linear function with slope $\frac{\pi}{4}$ (red line), which is due to the first law of entanglement entropy \cite{Bhattacharya:2012mi,Blanco:2013joa,Wong:2013gua} as
 will be analyzed in the next section. At larger $M$, this function provides an upper bound to the entanglement entropy, a consequence of the positivity of relative entropy \cite{Blanco:2013joa,Wong:2013gua}. Including constants $R$ and $G_N$, and using the dimension $\Delta$ of the local operator, this bound is expressed as
\begin{equation}
\label{EQ_EE_T0_BOUND}
\Delta S_A(t=0) \leq \frac{\pi M}{4 G_N R} = \frac{\pi}{2} R m = \frac{\pi}{2} \Delta \text{ .}
\end{equation}
We will explore this bound in more detail in the next section.

By inspecting the numerical solutions in figure \ref{FIG_T0_SURFACES}, we can also obtain an analytical form of $\Delta S_A(t=0)$ in the large $M$ limit. As the minimal surface wraps closer around the horizon with increasing $M$, we can approximate its shape by a half-sphere around the horizon.  Thus, we expect an approximate behavior
\begin{equation}
\Delta S_A(t=0) \simeq \frac{A_\text{hor}/2}{4 G_N} = \frac{\pi M^{2/3}}{2 G_N} \text{ ,}
\end{equation}
where we inserted the area $A_\text{hor}$ of the horizon at radius $r_\text{hor} \simeq M^{1/3}$. As we can see in figure \ref{FIG_EE_T0}, our approximation (orange line) is valid at large $M$, as expected.\footnote{
We ignored terms from the annulus region around the half-sphere, as well as the $M=0$ contribution to $\Delta S_A$. Together, these add a subleading contribution of $-\frac{3\pi}{8} M^{1/3}$.} For comparison, consider the exact  AdS$_3$ result \ref{EQ_ADS3_DELTA_SA_T0} (dashed line in figure \ref{FIG_EE_T0}). Expanding it in powers of $M$ yields a similar expression, where $A_\text{hor}$ is replaced by the horizon diameter $d_\text{hor} = 2 \pi M^{1/2}$.\footnote{
Note that in terms of physical mass $m$, the parameter $M$ varies with dimension $d$ according to \ref{EQ_ADS_MASS}.} This suggests that in AdS$_{d+1}$/CFT$_d$, the initial half-space entanglement entropy of a local quench at large $M$ can be written as
\begin{equation}
\Delta S_A^{\text{CFT}_d}(t=0) \simeq \frac{A^{S_d}_\text{hor}/2}{4 G_N} = \frac{\pi^{d/2} M^{(d-1)/d}}{4 G_N \Gamma(\frac{d}{2})} \text{ .}
\end{equation}

\begin{figure}[ht!]
\centering
\includegraphics[width=0.52\textwidth]{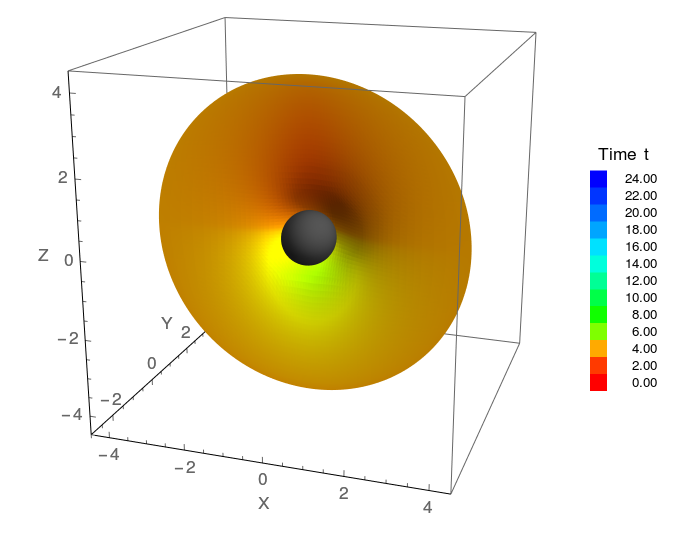}
\includegraphics[width=0.52\textwidth]{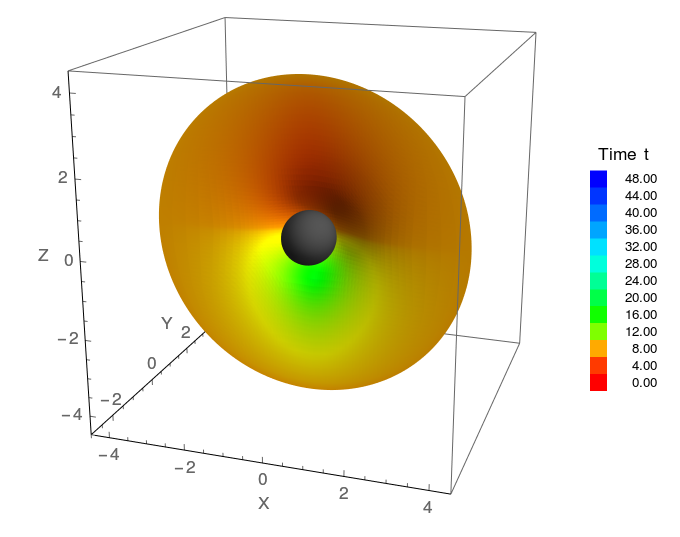}
\includegraphics[width=0.52\textwidth]{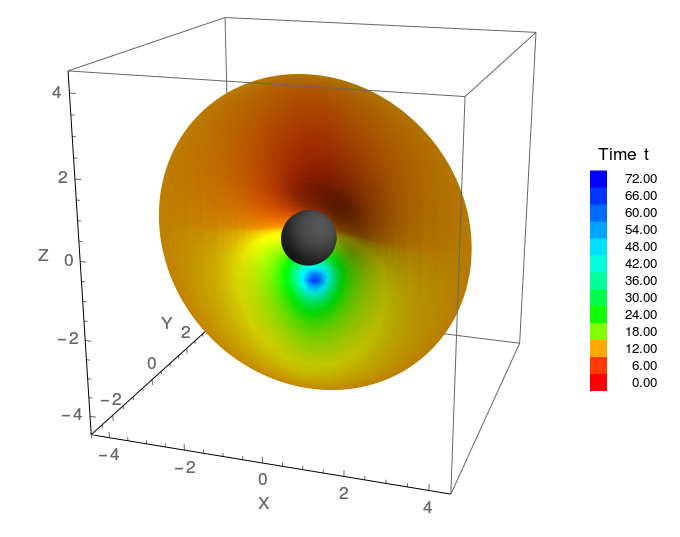}
\caption{Numerically extremized surfaces in global coordinates for mass parameter $M=1$ at quench time $t=4$ (top), $t=8$ (middle), and $t=12$ (bottom). The black spheres show the horizon of the metric. Coordinate labels are $(X,Y,Z)=(r \sin\theta \cos\phi,r \sin\theta \sin\phi,r \cos\theta)$ with the time coordinate color-coded relative to $t$ at the boundary. The original computation was performed with a global time coordinate, converted here to Poincaré time $t$. Units in $R=\alpha=1$.}
\label{FIG_T_VARIED_SURFACES}
\end{figure}

For the $t>0$ case, we need to compute extremal surfaces in the full 4-dimensional spacetime. While boundary time $t$ is constant, it may vary on the rest of the surface. In order to visualize the solutions, we color-code vertices in their local time coordinate. The change in shape from the $t=0$ solution is shown in figure \ref{FIG_T_VARIED_SURFACES} for $M=1$. The surfaces are no longer isotropic and the local surface time differs considerably from boundary time $t$. Around $\cos\theta<0$, which corresponds to the $z<\sqrt{\alpha^2 + t^2}$ region in Poincaré coordinates between falling mass and AdS boundary, the surface dips into the local future. It returns to boundary time $t$ at its closest distance to the mass, and then dips into the past around $\cos\theta>0$.

\begin{figure}[htb!]
\centering
\includegraphics[width=0.63\textwidth]{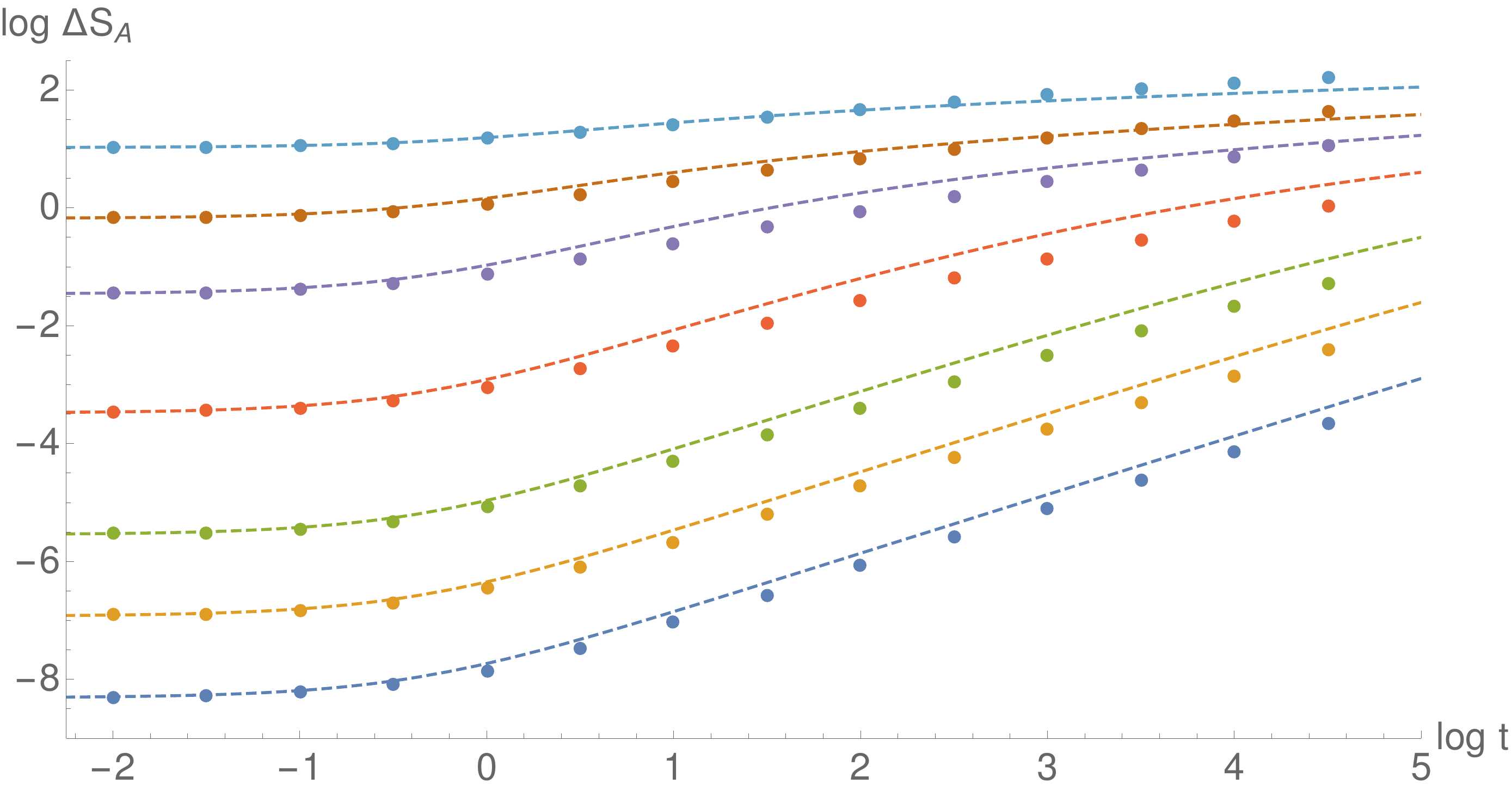}
\caption{Time dependence of the growth of entanglement entropy $\Delta S_A$ after a local quench. With mass parameters $M=0.01\cdot2^n$, with $n=9$, 7, 5, 2, -1, -3 and -5 (from top to bottom). Data points show numerical computations in AdS$_4$, dashed lines are the analytical AdS$_3$ result rescaled to match the AdS$_4$ data at $t=0$. Units in $G_N = R = \alpha = 1$.}
\label{FIG_EE_M_T_VARIED}
\end{figure}

\begin{figure}[htb!]
\centering
\includegraphics[width=0.63\textwidth]{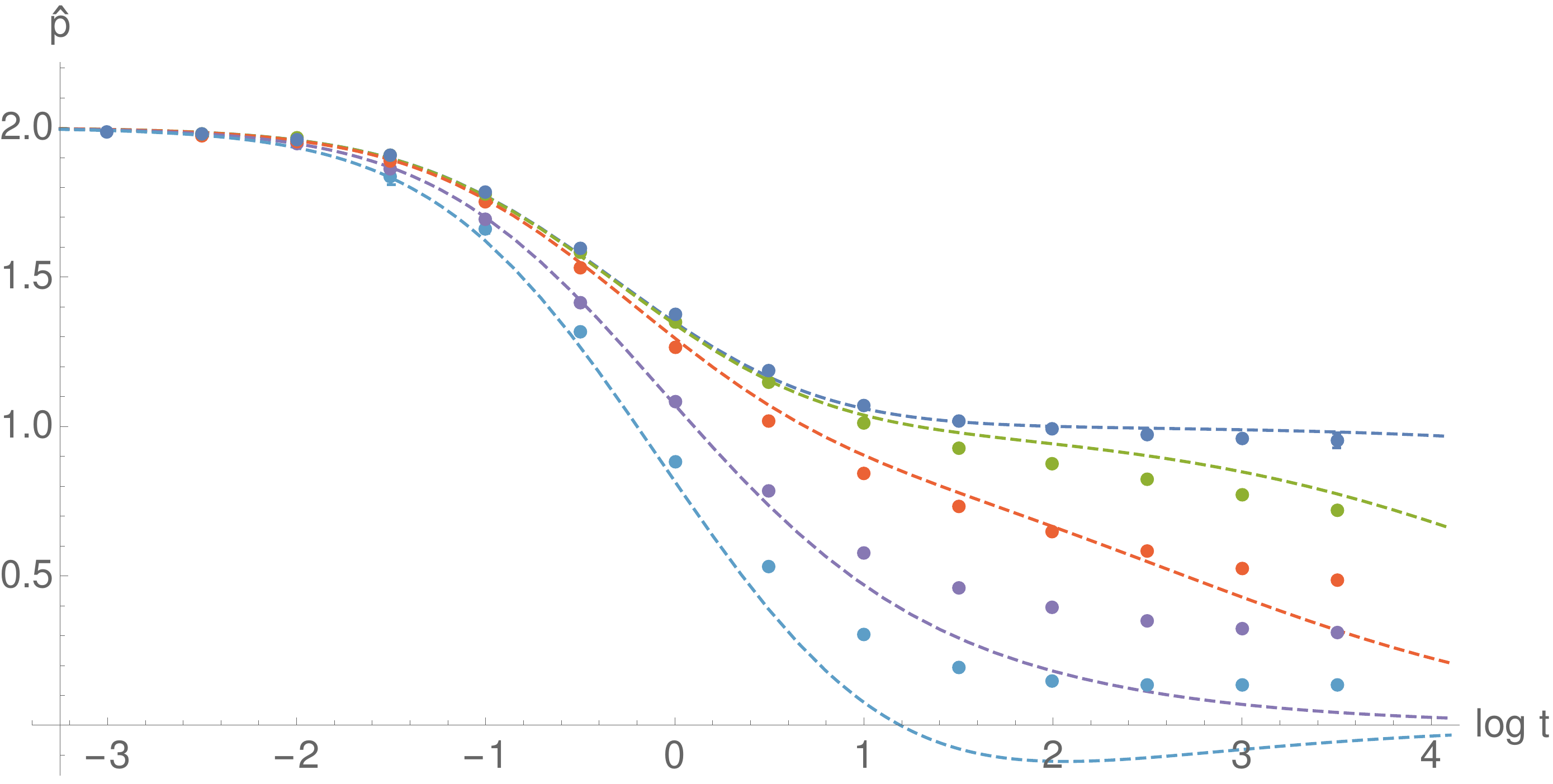}
\caption{Local power coefficient of $\Delta S_A$ for mass parameters $M=0.01\cdot2^n$, with $n=9$, 5, 2, -1, and -5 (from bottom to top). Data points show numerical computations in AdS$_4$, dashed lines are the analytical AdS$_3$ result (no rescaling). Units in $G_N = R = \alpha = 1$.}
\label{FIG_EE_M_T_VARIED_P}
\end{figure}

\begin{figure}[htb!]
\centering
\includegraphics[width=0.63\textwidth]{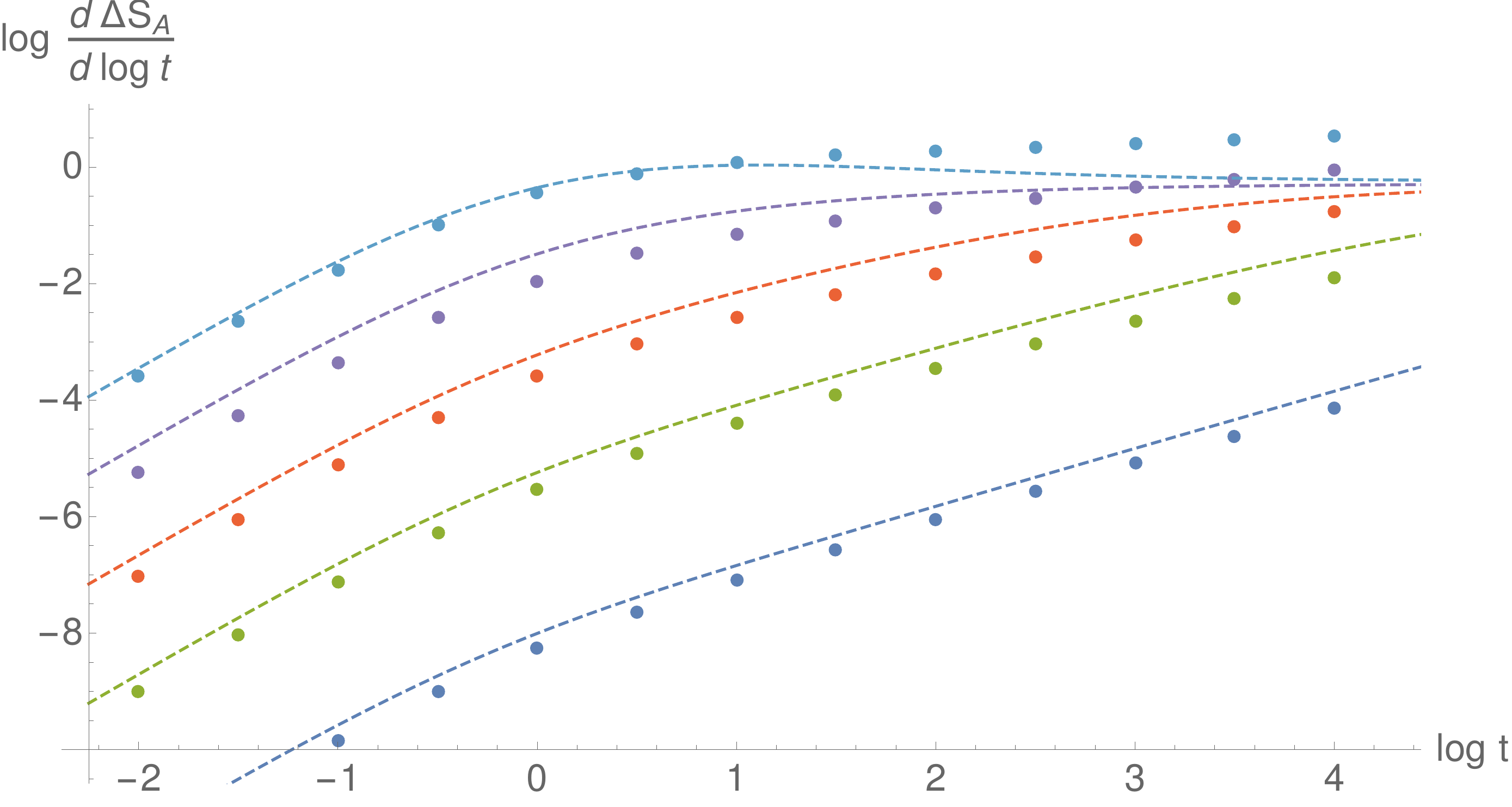}
\caption{Local logarithmic coefficient of $\Delta S_A$ for the same data set as in figure \ref{FIG_EE_M_T_VARIED_P}, with $M$ now decreasing from top to bottom. The analytical AdS$_3$ result (dashed lines) has been rescaled to match the AdS$_4$ data at $t=0$. Units in $G_N = R = \alpha = 1$.}
\label{FIG_EE_M_T_VARIED_EE_LOG_DSA}
\end{figure}

With increasing time $t$ at the boundary, the surface itself becomes more time-dependent. For large $t$ and $M$, parts of the extremal surface cross the horizon of the Poincaré coordinate patch, forcing us to perform all computations in global coordinates.

The numerical results for the time evolution of $\Delta S_A$ are shown in figure \ref{FIG_EE_M_T_VARIED} for a range of mass parameters $M$. The exact result for the AdS$_3$/CFT$_2$ case (dashed lines), given by \eqref{EQ_ADS3_DELTA_SA_GENERAL}, is shown for the same values of $M$. For easier comparison, it has been rescaled to match the AdS$_4$/CFT$_3$ data at $t=0$. We find clear deviations between them, especially at late time.

We consider two possible scenarios at large $t$: An asymptotic power law $\Delta S_A(t)\propto
t^{p}$, and a logarithmic time dependence $\Delta S_A(t)\propto \log t$, the latter of which we observed in the AdS$_3$ case.

To test for an asymptotic power law, we use an estimator $\hat{p}$ of the power coefficient $p$ of a supposed asymptotic time dependence $f(t) = a + b \,t^p$. Assume a triplet of data points $\left((e^{t_0- \Delta t}, f_1),\; (e^{t_0}, f_2),\; (e^{t_0 + \Delta t}, f_3) \right)$ is given. If $f(t)$ matches these data points, the estimated power coefficent follows as
\begin{equation}
\label{EQ_POWER_COEFFICIENT}
\hat{p} = \frac{1}{\Delta t} \log\left( \frac{f_3-f_2}{f_2-f_1} \right) \text{ .}
\end{equation}
By Gaussian error propagation, the associated absolute error is
\begin{equation}
\delta \hat{p} = \frac{1}{\Delta t} \sqrt{\frac{(f_3-f_2)^2 \delta f_1^2 + (f_3-f_1)^2 \delta f_2^2 + (f_2-f_1)^2 \delta f_3^2}{(f_3-f_2)^2 (f_2 - f_1)^2}}
\end{equation}
in terms of the absolute errors $\delta f_k$ of the function values. Estimators for $a$ and $b$ can be constructed analogously.

The results for $f(t) = \Delta S_A(t)$ are shown in figure \ref{FIG_EE_M_T_VARIED_P}. In the limit $t \to 0$, the power coefficient converges to $\hat{p}=2$ for all $M$. For $t \to \infty$, however, $\hat{p}$ strongly depends on $M$: At small $M$, an apparently unstable plateau close to $\hat{p}=1$ is visible, while at large $M$, there appears to be convergence to a small but nonzero power.
At large $t$, a strict bound $\hat{p} < 1$ is satisfied, i.e.\ the entanglement entropy increases sub-linearly. Note that the initial $\hat{p}=2$ growth and $\hat{p}=1$ plateau follow from the first law of entanglement entropy, to be explained in the next section. The late time sub-linear
growth is a genuinely new behavior, which we present in this paper for the first time.

In order to test a possible logarithmic time dependence $\Delta S_A(t)=a + b\, \log{t}$ at large $t$, we simply check $\frac{\text{d}\,\Delta S_A(t)}{\text{d}\log{t}}$ for convergence. This is plotted in figure \ref{FIG_EE_M_T_VARIED_EE_LOG_DSA}. While the AdS$_3$ result converges to a constant, the numerical data for the AdS$_4$ case shows no sign of slowing down within the range of $t$ that we computed.

Unfortunately, our numerical approach cannot be extended to arbitrarily large $t$, as the increasing time dependence requires a larger finite element resolution, slowing down the algorithm considerably. Thus, we can cannot exclude the possibility that the apparent asymptotic power law breaks down at very large $t$, which might lead to a logarithmic growth in the end.
However, our AdS$_4/$CFT$_3$ analysis shows clear differences to the AdS$_3/$CFT$_2$ case, which extend to at least $\frac{t}{\alpha} \approx 50$. 

In fact, there are a few theoretical hints suggestive of an asymptotically logarithmic time evolution $\Delta S_A(t)\propto \log t$. For instance, in an analysis of Renyi entanglement entropy for local quenches of holographic CFTs we find a logarithmic time evolution \cite{Caputa:2014vaa} in any dimensions $d$ when $\Delta\ll c_{d}$, where $c_d=O(N^2)$ describes the degrees of freedom of the CFT. However, this analysis breaks down in the von Neumann entropy limit $n=1$. Moreover, the tensor network description of a falling particle proposed in \cite{Nozaki:2013wia} for 2-dimensional CFTs is straightworward to generalize to any dimensions, also leading to a logarithmic growth of entanglement entropy.

In contrast, the theoretical background of the initial time evolution leads to much less ambiguity, as our observations fully agree with the first law of entanglement entropy.

\section{Local Quenches and the First Law of Entanglement Entropy}
\label{S_4}

We shall now give an interpretation of some parts of our numerical results presented in the previous section. Consider the relative entropy $S(\rho_1 | \rho_2)$ between two different quantum states, defined as
\begin{equation}
S(\rho_1 | \rho_2) = \tr{\rho_1 \log\rho_1} - \tr{\rho_1 \log\rho_0} \text{ ,}
\end{equation}
where $\rho_i$ is the density matrix corresponding to the $i$th state. The positivity of  $S(\rho_1 | \rho_2)$ leads to a bound to the entanglement entropy
change \cite{Blanco:2013joa,Wong:2013gua}:
\be
S(\rho_1 | \rho_2) = \Delta \la H_A\lb - \Delta S_A \geq 0 \text{ ,} \label{first_ineq}
\ee
where $\Delta \la H_A\lb $ denotes the change in expectation value of the modular Hamiltonian $H_A=-\log\rho_A$. In particular, if we consider a $d$-dimensional CFT and choose the subsystem $A$ to be a round ball with radius $l$, we can explicitly write the change of $H_A$ between an excited states and the CFT vacuum as follows \cite{Blanco:2013joa,Wong:2013gua}:
\be
\Delta \la H_A\lb = 2\pi \int_{|x-x_0|\leq l} (dx)^{d-1} \left(\f{l^2-|x-x_0|^2}{2l}\right) T_{tt}(x) \text{ .}
\label{firstcft}
\ee
Here $x_0$ is the center of the round ball $A$ and $T_{tt}$ is the energy density.

If we consider an infinitesimally small excitation, then the leading linear order contribution saturates the inequality \eqref{first_ineq}, leading to 
\begin{equation}
\Delta S_A = \Delta \la H_A\lb \text{ .} \label{first}
\end{equation}
This is called the first law of entanglement entropy \cite{Bhattacharya:2012mi,Blanco:2013joa,Wong:2013gua}, relating a change in energy (or energy density) to a change in entanglement entropy.

\subsection{AdS$_3$/CFT$_2$}

The energy density of local quenches \cite{Nozaki:2013wia} can be obtained in the AdS$_3$ setup as
\be
T_{tt}=\frac{\ap^2 \Delta}{\pi}\cdot\left[\frac{1}{((x-t)^2+\ap^2)^2}
+\frac{1}{((x+t)^2+\ap^2)^2}\right] \text{ .}
\ee
The subsystem $A$ is taken as an interval $x \in [0,2l]$ and the local excitation is situated at $x=0$. The first law (\ref{first}) leads to
\ba
\Delta S_A&=&2\pi\cdot\frac{\ap^2 \Delta}{\pi}\cdot \int^{2l}_0 dx \frac{x(2l-x)}{2l}
\left[\frac{1}{((x-t)^2+\ap^2)^2}+\frac{1}{((x+t)^2+\ap^2)^2}\right]\no
&=&\frac{\Delta}{2 l\ap}\Biggl[(2lt-t^2-\ap^2)\arctan\left(\frac{2l-t}{\ap}\right)
+4l\left(\ap+t\arctan\left(\frac{t}{\ap}\right)\right)\no
&& \ \ \ \ \ \ \ \ \ -(2lt+t^2+\ap^2)
\arctan\left(\frac{2l+t}{\ap}\right)\Biggr] \text{ .}
\ea
We now take $l \to \infty$, extending $A$ to the entire half-line. We find
\ba
\Delta S_A = 2 \Delta \left( 1 + \frac{t}{\ap} \arctan \frac{t}{\ap} \right) =
  \begin{cases}
    2 \Delta \left(1 + \frac{t^2}{\ap^2} \right) & \quad \text{for } t\ll \ap \\
    \frac{\pi \Delta}{\alpha} t  & \quad \text{for } t\gg \ap \\
  \end{cases} \text{ .}\label{qq}
\ea
Using the identity \cite{Gubser:1998bc_Witten:1998qj}
\be
\Delta =mR=\frac{(d-1)\pi^{d/2-1} M}{8\Gamma(d/2) G_N R}\text{ ,}
\ee
we can rewrite this result for AdS$_3$/CFT$_2$ ($d=2$) in terms of the mass parameter $M$:
\ba
\Delta S_A =
  \begin{cases}
    \frac{M}{4 G_N R} \left(1 + \frac{t^2}{\ap^2} \right) & \quad \text{for } t\ll \ap \\
    \frac{\pi M}{8 G_N R \alpha} t  & \quad \text{for } t\gg \ap \\
  \end{cases} \text{ .}
\ea

Thus, the first law predicts quadratic growth of $\Delta S_A$ at early time and a transition to linear growth at later time. However, in order to apply the first law we require $\Delta S_A\ll c$, where $c$ is the central charge \cite{Bhattacharya:2012mi}. At small $t$, \eqref{qq} turns this into the condition $\Delta \ll c$. At larger $t$, we get the additional constraint
\be
\frac{t}{\ap} \ll \frac{c}{\Delta} = \frac{12 R^2}{M} \text{ ,} \label{cond}
\ee
where we used the holographic relation $\frac{\Delta}{c}=\frac{M}{12R^2}$. If both constraints are met, our results from the first law relations can be trusted. On the other hand, when  $\frac{t}{\ap} \gg \frac{c}{\Delta}$ (late time zone), we have the logarithmic grow $S_A\simeq \frac{c}{6}\log t$, derived by both holographic and field theoretic methods \cite{Nozaki:2013wia,Asplund:2014coa}.

The exact results for $\Delta S_A$ in AdS$_3$/CFT$_2$, along with the AdS$_4$/CFT$_3$ computations, are shown in figure \ref{FIG_EE_M_T_VARIED_P} in terms of a local power fit \eqref{EQ_POWER_COEFFICIENT} with power coefficent $\hat{p}$. For a small mass parameter $M$, the first law requirement $\Delta S_A \ll c$ holds and we observe a quadratic growth ($\hat{p}=2$) at early time. In the intermediate region $1 \ll \frac{t}{\alpha} \ll \frac{c}{\Delta}$ , the predicted transition to linear growth ($\hat{p}=1$) can be seen. With increasing $M$, the time dependence becomes logarithmic more quickly, with $\hat{p}$ converging to zero.

\subsection{AdS$_4$/CFT$_3$}

In the AdS$_4$/CFT$_3$ case ($d=3$) the energy density looks like
\begin{equation}
T_{t t} = \frac{2\ap^3 \Delta}{\pi}\frac{(\alpha^2+t^2+\rho^2)^2 + 2 t^2 \rho^2}{\left((\alpha^2 + t^2-\rho^2)^2 + 4 \alpha^2 \rho^2\right)^{5/2}} \text{ ,}
\end{equation}
with a radial coordinate $\rho=\s{x^2+y^2}$. First, consider the configuration at small time $t \ll \ap$, where the energy density reads
\be
\label{EQ_T_TT_SERIES_ADS4}
T_{tt}=\frac{2\ap^3\Delta}{\pi} \left( \frac{1}{(\rho^2+\ap^2)^3} + \frac{9\rho^2 - 3 \ap^2}{(\rho^2 + \ap^2)^5} t^2 + O(t^4) \right) \text{ .}
\ee
We choose the subsystem $A$ to be a disk with a radius $l$ whose center is at the point $(x,y)=(l,0)$.
At $t=0$, the change of the modular Hamiltonian is
\ba
\label{EQ_D_HA_DERIVATION_ADS4}
\Delta\la H_A\lb &=&2\pi \int^l_0 \text{d}r\; r \int^{2\pi}_0 \text{d}\theta\; T_{tt}\cdot \frac{l^2-r^2}{2l},\no
&=& \frac{2\Delta\ap^3}{l} \int^l_0 \text{d}r \int^{2\pi}_0 \text{d}\theta\; \frac{r(l^2-r^2)}{(l^2+r^2+2lr\cos\theta+\ap^2)^3},\no
&=& \frac{4\pi\Delta\ap^3}{l}\int^l_{0}\text{d}r\; \frac{r(l^2-r^2) \left( (l^2 + r^2 + \ap^2)^2 + 2 l^2 r^2 \right)}{\left( (l^2 + r^2 + \ap^2)^2 - 4 l^2 r^2 \right)^{5/2}},\no
&\simeq & \frac{\pi}{2}\Delta \text{ ,}
\ea
where in the final expression we used $l \to \infty$. This leads to the  bound
\be
\Delta S_A\leq \frac{\pi \Delta}{2}=\frac{\pi M}{4G_N R} \text{ ,}
\ee
which reproduces the bound \eqref{EQ_EE_T0_BOUND} that we observed in figure \ref{FIG_EE_T0}.

The calculation steps in \eqref{EQ_D_HA_DERIVATION_ADS4} can extended to the $O(t^2)$ term in \eqref{EQ_T_TT_SERIES_ADS4}, yielding
\ba
\label{EQ_SA_SMALL_T_BOUND_ADS4}
\Delta S_A \leq \frac{\pi \Delta}{2} + \frac{3 \pi \Delta t^2}{8 \ap^2} + O(t^4) = \frac{\pi M}{4 G_N R} \left( 1 + \frac{3 t^2}{4 \ap^2} + O(t^4) \right)
\ea
as the bound to entanglement entropy growth at small $t$.

Next we assume $\ap\ll t \ll l$. Then the integral of the first law (\ref{first}) has a dominant contribution around $\rho\simeq t$. Thus it is approximated as follows:
\begin{alignat}{2}
\Delta \la H_A\lb &\simeq 2\pi\cdot \frac{\Delta \ap^3}{4\pi}\cdot \int^\infty_0 \text{d}\rho \int^\f{\pi}{2}_{-\f{\pi}{2}} & \text{d}\theta \left( \f{2l\cdot (l-\sqrt{(l-\rho\cos\theta)^2+t^2\sin^2\theta})}{2l} \right. \no
& &\quad \left. \cdot \f{3}{((\rho-t)^2+\ap^2)^{5/2}\s{4t^2+\ap^2}} \right) \no
& \simeq \f{\Delta}{\ap}t\cdot \int^\f{\pi}{2}_{-\f{\pi}{2}} \text{d}\theta \cos\theta \no
& =\frac{2\Delta}{\ap}t \text{ .}
\end{alignat}
Hence we find
\be
\label{EQ_SA_BOUND_ADS4}
\Delta S_A \leq \frac{2\Delta}{\ap}t=\frac{Mt}{G_N R\ap} \text{ .}
\ee
The saturation of this inequality (i.e.\ the first law) reproduces the perturbative holographic result (\ref{lqxx}). This is to be expected, as both the first law and the perturbative approach are valid in the regime of small excitations, i.e.\ when \eqref{first_ineq} only contains subleading contributions with respect to changes in the density matrix describing the system. In our setup, this corresponds to small $M$, where the shape of the extremal surface is not greatly perturbed by the mass. Under these conditions, we expect the qualitative time dependence of $\Delta S_A$ at early and intermediate time $t$ to be equivalent to the AdS$_3$/CFT$_2$ case, i.e.\ with an initial quadratic growth at $\frac{t}{\ap} \ll 1$ and a linear behavior at intermediate time. Indeed, that is what we already saw for the numerical data in figure \ref{FIG_EE_M_T_VARIED_P}.

\begin{figure}[tb!]
\centering
\includegraphics[width=0.63\textwidth]{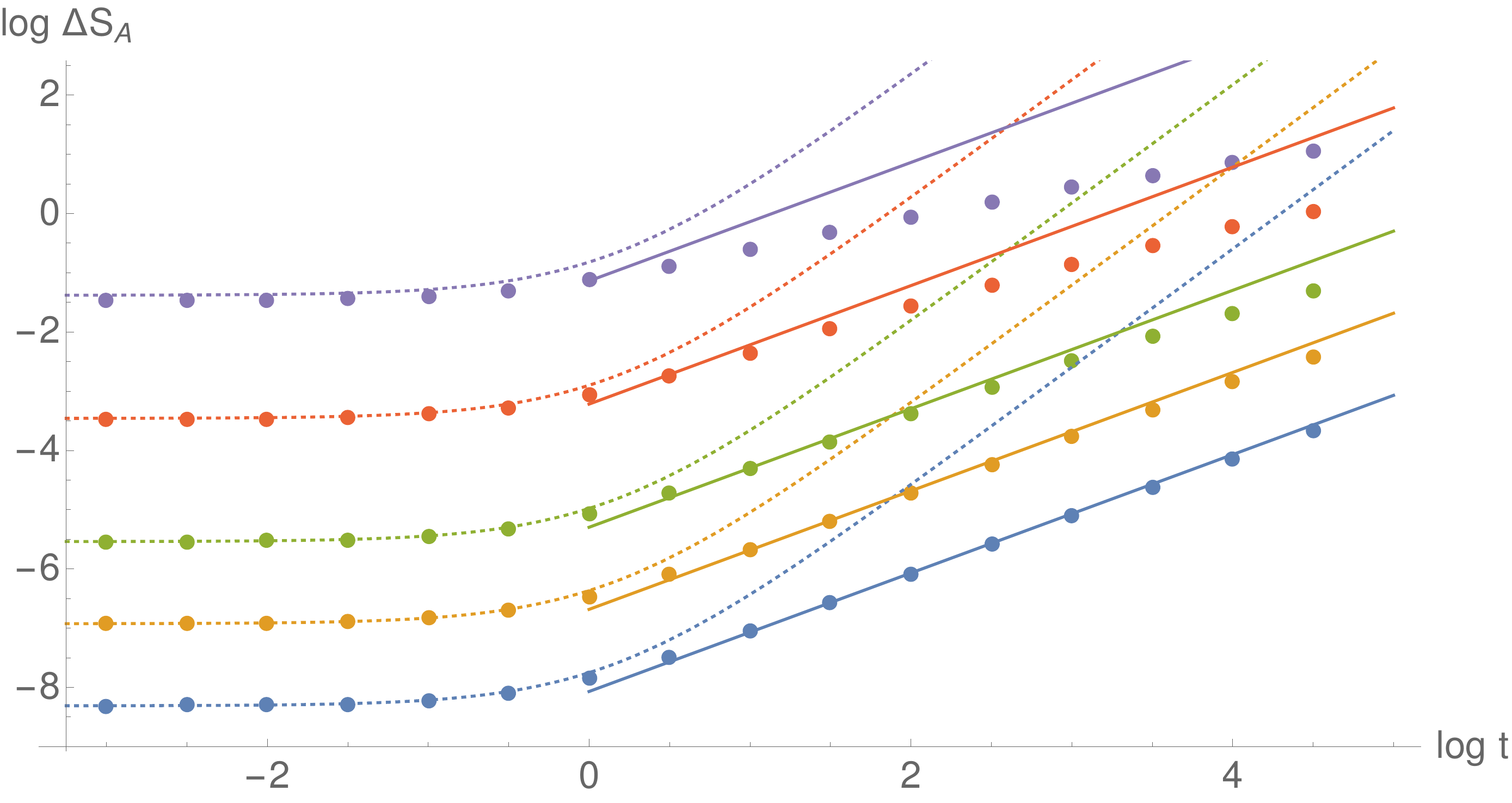}
\caption{Excited entanglement entropy $\Delta S_A$ for mass parameters $M=0.01\cdot2^n$, with $n=5$, 2, -1, -3, and -5 (from top to bottom). Data points show numerical computations in AdS$_4$, dotted lines show bound \eqref{EQ_SA_SMALL_T_BOUND_ADS4}, solid lines correspond to bound \eqref{EQ_SA_BOUND_ADS4}. Units in $G_N = R = \ap = 1$.}
\label{FIG_EE_BOUND}
\end{figure}

Even at $t=0$, the first law bound is not reliable if $\Delta$ is large. Thus, the behavior
\begin{equation}
\Delta S_A(t=0) \simeq \frac{\pi M^{2/3}}{2 G_N} = \frac{\pi (R \Delta)^{2/3}}{(2 G_N)^{1/3}}
\end{equation}
that we observed at large $M$ in figure \ref{FIG_EE_T0} cannot be understood in terms of the first law.

The time-dependent bounds \eqref{EQ_SA_SMALL_T_BOUND_ADS4} and \eqref{EQ_SA_BOUND_ADS4} (for  $t \ll \ap$ and $t \gg \ap$, respectively) are shown in figure \ref{FIG_EE_BOUND} alongside numerical data at small $M$. In their respective $t$ range, they provide an upper bound for all $M$ and saturate as $M \to 0$, as expected. As in the AdS$_3$/CFT$_2$ case, we expect the saturation to break down at very large $t$ as the entanglement entropy reaches its asymptotic, sub-linear growth.

\newpage
\section{Summary and Discussion}

In this work, we studied the holographic entanglement entropy of an AdS$_4$/CFT$_3$ setup involving a local quench. To this end, we employed a numerical finite-element approach to calculate extremal surfaces in the AdS$_4$ geometry dual to the CFT$_3$. In particular, we considered the growth of entanglement entropy $\Delta S_A$ corresponding to a half-space subsystem $A$, on the boundary of which the local quench occurs. The strength of the quench is determined by the operator dimension $\Delta$ of the excitation, which is proportional to a mass $m$ that is falling freely in the bulk theory \cite{Nozaki:2013wia}.

At time $t=0$ after the quench, the numerical data (see figure \ref{FIG_EE_T0}) is constrained by a bound
\begin{equation}
\Delta S_A(t=0) \leq \frac{\pi \Delta}{2}
\end{equation}
which saturates at small $\Delta$, in agreement with the first law of entanglement entropy. At large $\Delta$, the extremal surface area is dominated by the region near the coordinate horizon around the mass, leading to the relation
\begin{equation}
\Delta S_A(t=0) \simeq \frac{A_\text{hor}/2}{4 G_N} = \frac{\pi (R \Delta)^\frac{2}{3}}{(2 G_N)^\frac{1}{3}} \text{ ,}
\end{equation}
where $A_\text{hor}$ is the horizon area and $R$ the AdS radius. As a similar relationship can be derived for the AdS$_3$/CFT$_2$ case, a relation
\begin{equation}
\Delta S_A^{\text{CFT}_d} ~ (t=0) \simeq \frac{A_\text{hor}^{S_d}/2}{4 G_N} = \frac{2 \pi^\frac{3}{2}}{(8 \pi G_N \Gamma(\frac{d}{2}))^{\frac{1}{d}}} \left(\frac{R \Delta}{d-1} \right)^{1-\frac{1}{d}}
\end{equation}
can be obtained for local quenches in holographic $d$-dimensional CFT in the large $\Delta$ limit.

At small $t$ and $\Delta$, our numerical results exhibit an initial quadratic bound that is again in agreement with the first law prediction
\ba
\Delta S_A \leq \frac{\pi \Delta}{2} + \frac{3 \pi \Delta t^2}{8 \ap^2} + O(t^4) \text{ .}
\ea
At larger $t$, this turns into a bound linear in $t$:
\be
\Delta S_A \leq \frac{2\Delta}{\ap}t \text{ .}
\ee
Again, both bounds are saturated at small $\Delta$ (see in figure \ref{FIG_EE_BOUND}).

Qualitative differences to the entanglement growth in 2-dimensional CFT become apparent at large $t$ and $\Delta$: Instead of quickly reaching an asymptotically logarithmic time dependence, we observe a growth similar to a power law $\Delta S_A\sim t^{p}$ (see figure \ref{FIG_EE_M_T_VARIED_P} and \ref{FIG_EE_M_T_VARIED_EE_LOG_DSA}) in the late time region of our analysis.

This observed time dependence may not be truly asymptotical, as our numerical methods cannot be extended to arbitrarily large time $t$. However, it is clear that significant deviations from the entanglement entropy in lower dimensions appear with increasing $t$. It is also interesting to note that the power $p$ (shown by an estimator $\hat{p}$ in figure \ref{FIG_EE_M_T_VARIED_P}) monotonically decreases toward zero as the time $t$ and the mass parameter $M$ (and thus $\Delta$) get larger. This suggests that the asymptotic limit is either $\Delta$-dependent, or slower than a power law, i.e.\ logarithmic in $t$. The latter case would be more consistent with arguments from Renyi entropies \cite{Caputa:2014vaa} and tensor networks \cite{Nozaki:2013wia}, but clearly, complete analytical studies of entanglement entropy evolution in higher-dimensional CFTs will be needed to explain our results. This would require further characterization of holographic CFTs in the language of field theory.

We would also like to mention that in our holographic approach \cite{Nozaki:2013wia}, we regard the gravity dual of local quench as a massive heavy particle falling in the bulk AdS. This treatment precisely corresponds to the large $c$ approximation \cite{Asplund:2014coa}. However, more precisely it should be described by a time-dependent classical solution made of a scalar field in the bulk AdS with gravitational backreactions as in \cite{Rangamani:2015agy}. Such a detailed structure of massive excitation in the bulk may affect the late time evolution of holographic entanglement entropy, which is dual to the breakdown of the standard large $c$ approximation. It is a very interesting future problem to take this effect into account to see if the entanglement entropy in the late time limit can approach a finite constant or continue to grow forever.

In principle, the finite-element optimization strategy used in this paper can be applied to a large class of problems involving holographic entanglement entropy in AdS$_4$/CFT$_3$. While our current implementation (details in appendix \ref{APP_A}) only considers simply connected boundary regions, extensions to more complicated models are possible. As analytical calculations in CFT$_3$ are notoriously complicated, our numerical technique can offer an alternative approach.

\subsection*{Acknowledgements}
%%%%%%%%%%%%%%%%%%%%%%%%%%%%%%%

We would like to thank  Vijay Balasubramanian, David Berenstein, Pawel Caputa, Jens Eisert, Mario Flory, Masairo Nozaki, Tokiro Numasawa, Erik Tonni and Herman Verlinde for useful discussions, and especially to Robert Myers for insightful comments.
 TT is supported by the Simons Foundation through the ``It from Qubit'' collaboration and JSPS Grant-in-Aid for Scientific Research (A) No.\ 16H02182. TT is also supported by World Premier International Research Center Initiative (WPI Initiative) from the Japan Ministry of Education, Culture, Sports, Science and Technology (MEXT). AJ is supported by the German Academic Scholarship Foundation (\textsl{Studienstiftung des deutschen Volkes}). His research exchange to Kyoto University was supported by a grant of the German Academic Exchange Service (\textsl{DAAD}). AJ is also grateful to Valentina Forini for supporting the thesis on which this work is partly based. We are very grateful to the long term workshop “Quantum Information in String Theory and Many-body Systems” held at YITP in Kyoto University and the IGST 2016 conference held at Humboldt University, where this work was partially conducted.

\newpage
\thispagestyle{empty}
\appendix
\section{Finite Element Implementation}
\label{APP_A}
This appendix covers the main features of our algorithm for finding the area of an extremal 2-surface with a given boundary in 3+1 spacetime dimensions. We implemented the algorithm in C\texttt{++} without the use of any existing framework. Visualizations of the output surface data were produced with \textsl{Mathematica}.

We use a nomenclature customary in graphics programming: A discretization point is called a \textsl{vertex}, a line segment between two vertices is referred to as an \textsl{edge}, and a surface element is a \textsl{face}. Faces are typically triangles with three corner vertices or quadrilaterals, also called \textsl{quads}, with four. The full geometry is called a \textsl{mesh}.

In terms of data structures, a vertex is a collection of four floating-point numbers, one for each coordinate. An edge, a triangle, and a quadrilateral are collections of two, three, and four integers, respectively. Each integer serves as an index pointing to a vertex. Thus, only the vertices are dynamical objects, while edges and faces are derived objects used to calculate lengths and areas of the geometry.

The type of numerical approach used here is a \textsl{finite element method}, where a continuous problem is solved on a discretized mesh (for an introduction, see e.g.\ \cite{Zienkiewicz1977,Hughes2000}). Shape optimization problems of this type are often encountered in engineering, as in the minimization of the strain on a component. In our case, the functional to be optimized is the area of the surface itself. However, extending this approach to surfaces in 4-dimensional curved spacetime leads to several complications, described throughout the rest of this appendix.

\subsection{Area calculation}
The primitive surface element for computing areas is the triangle. A triangle spanned by two differential vectors $\text{d}v^\mu$ and $\text{d}w^\mu$ on a Riemannian manifold with metric $g_{\mu \nu}$, has an area of
\begin{align}
\label{EQ_TRI_DAREA1}
\text{d}A_\bigtriangleup = \frac{1}{2} | \text{d}v \land \text{d}w | &=  \frac{1}{2} \sqrt{|\text{d}v|^2 |\text{d}w|^2 - |\text{d}v \cdot \text{d}w|^2} \nonumber \\
&= \frac{1}{2} \sqrt{(g_{\mu \nu} \text{d}v^\mu \text{d}v^\nu) (g_{\rho \sigma} \text{d}w^\rho \text{d}w^\sigma) - (g_{\mu \nu} \text{d}v^\mu \text{d}w^\nu)^2} \text{ ,}
\end{align}
where we used Lagrange's identity for the wedge product $a \land b$ to get to the second step. If we parametrize the differential vectors as the derivative of a function $x^\mu$ along two coordinates $t$ and $u$, the result is equivalent to half of the familiar differential area element:
\begin{align}
\label{EQ_TRI_DAREA2}
\text{d}A_\bigtriangleup &= \frac{1}{2} \sqrt{\left(g_{\mu \nu} \frac{\text{d}x}{\text{d}u}^\mu \frac{\text{d}x}{\text{d}u}^\nu\right) \left(g_{\rho \sigma} \frac{\text{d}x}{\text{d}v}^\rho \frac{\text{d}x}{\text{d}v}^\sigma\right) - \left(g_{\mu \nu} \frac{\text{d}x}{\text{d}u}^\mu \frac{\text{d}x}{\text{d}v}^\nu\right)^2} \; \text{d}u\, \text{d}v \nonumber \\
&= \frac{1}{2} \sqrt{\det G_{\mu \nu}} \; \text{d}u\, \text{d}v
\text{ ,}
\end{align}
where $G_{\mu \nu}$ is the induced metric on the surface parameterized by $x^\mu(u,v)$. In principle, we could integrate \eqref{EQ_TRI_DAREA2} over $u$ and $v$ to yield the exact area formula of a triangle in a spacetime with metric $g_{\mu \nu}$. However, this metric is usually too complicated to yield a compact analytical expression. Instead, we can assume that each face is small enough so that $g_{\mu \nu}$ varies little along its surface. Then we simply use \eqref{EQ_TRI_DAREA1} with $\text{d}v^\mu \to p_1^\mu-p_3^\mu$ and  $\text{d}w^\nu \to p_2^\nu-p_3^\nu$ as the formula for the area of a triangle with corner vertices $p_1$, $p_2$ and $p_3$. As we subdivide the mesh into smaller and smaller elements, the error associated with this approximation gradually decreases.

Clearly, an area can only be defined for a space-like triangle, i.e.\ one for which \eqref{EQ_TRI_DAREA1} is real. However, the parameter space of all dynamic vertices allows for unphysical, time-like solutions. Approximating edges as differential vectors can also lead to configurations where \eqref{EQ_TRI_DAREA1} turns imaginary. While choosing a coordinate system that minimizes time dependence alleviates this problem (see \ref{ss_timeslice_constraints}), it still forces us to use an optimization algorithm that avoids locally time-like solutions (outlined in \ref{ss_numerical_optimization}). First, however, we have to create a triangular discretization of the mesh we seek to extremize.

\subsection{Triangulation and quadrilateralization}

The boundary of the extremal surface is given as a closed chain of static vertices, chosen to approximate a continuous boundary function, between which we construct an initial mesh. After adding an additional vertex in the center of the boundary vertices and connecting every second boundary vertex with the center via an edge, we can define quads that produce a complete mesh.

The algorithm proceeds through $N_i$ iterations of first optimizing the mesh at a given discretization level and then refining it by subdividing the mesh into smaller quads. In order to calculate the area of a quad using \eqref{EQ_TRI_DAREA1}, we need to divide it into a pair of triangles. However, depending on which quad diagonal is chosen as the separation between both triangles, there are two distinct ways of filling a quad. As the total surface has an orientation, these two solutions correspond to making the surface either locally convex or concave. As we subdivide the mesh into ever smaller quads and converge to a smooth surface, both solutions become equal. Therefore, we can choose the area of each quad as the average of the area of its two triangular fillings. From the perspective of the entire mesh, we average between an ``outer'' and ``inner'' triangulation that become more degenerate with each iteration step.

\begin{figure}[t]
\centering
\includegraphics[width=0.6\textwidth]{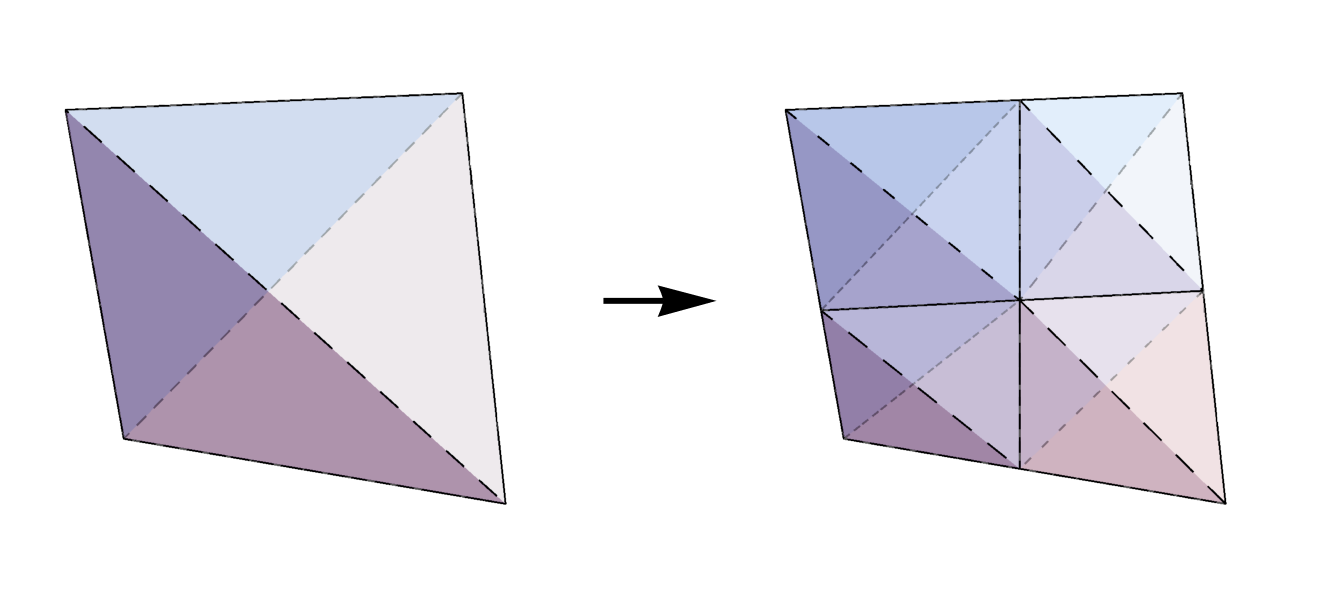}
\caption{Subdivision of a quad into four smaller quads. Both concave and convex triangular fillings are shown. Edges of quads are drawn as solid lines, while triangular edges are dashed.}
\label{FIG_TRIANGULATION}
\end{figure}

\begin{figure}[t]
\centering
\includegraphics[width=\textwidth]{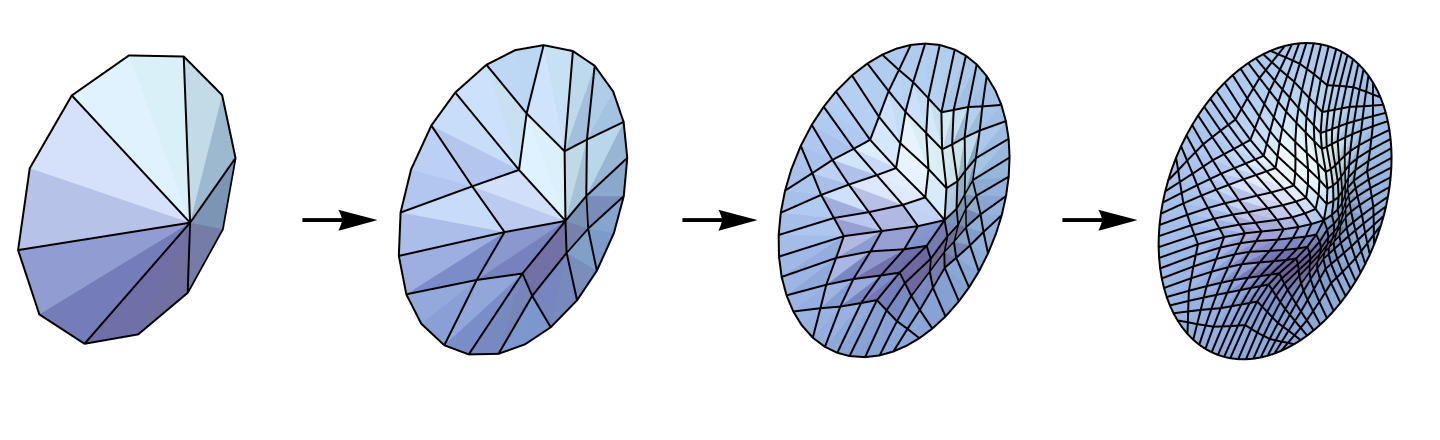}
\caption{Recursive subdivision of a quadrilateralized mesh. At each step, the mesh is extremized within the background metric (in this example, an AdS$_4$ black hole metric).}
\label{FIG_SUBDIVISION}
\end{figure}

The subdivision process of an individual quad is visualized in figure \ref{FIG_TRIANGULATION}. Each edge is divided into two, and an additional vertex is added in the center. Thus, each quad is divided into four smaller quads. The position of the new vertices is given by a weighted average of two (edge vertices) or four (center vertex) 4-vectors of the old vertices. The choice of weights depends on the problem: For calculating absolute areas, subdivision vertices should be chosen so that the new quads have roughly equal area. For the calculation of relative areas (i.e.\ between different metrics for similar boundary conditions), faster convergence can be reached by associating a greater weight to regions were deviations between both problems are largest.

Note than for the purposes of area calculation and subdivision, we treat edges as straight lines, a notion that depends on the choice of coordinates. Working with coordinate-independent geodesics is much more computationally demanding, if geodesic solutions are not explicitly known. As the edge length decreases exponentially with the number of iterations, the discrepancy quickly vanishes as long as the solution is sufficently smooth.

A full example of the quadrilateralization of a mesh with 12 initial boundary vertices is shown in figure \ref{FIG_SUBDIVISION}. Note that boundary edges are subdivided so that the new vertices follow the continuous boundary function.

\begin{figure}[t]
\centering
\includegraphics[width=0.7\textwidth]{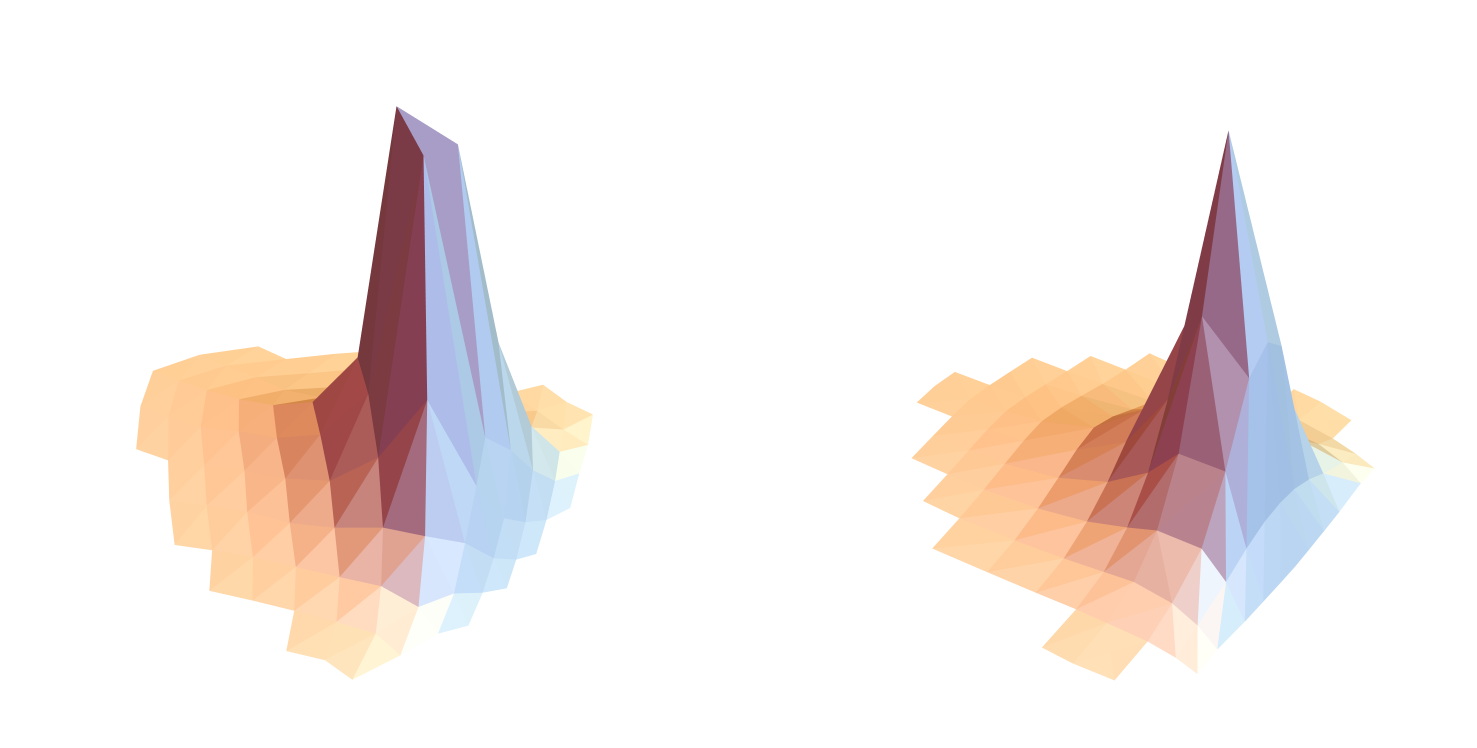}
\caption{Triangulation (left) and quadrilateralization (right) around a sharply peaked solution. Quads are shown as convex pairs of triangles. In both cases, vertices are at local optima with respect to variations along the vertical direction.}
\label{FIG_QUADS_VS_TRIS}
\end{figure}

The use of a quadrilateralized mesh instead of a simpler triangulated one is preferable due to its stability with respect to subdivision. As quads are locally agnostic with respect to concavity, the optimization algorithm cannot get trapped in a local concave minimum. The problem is visualized in figure \ref{FIG_QUADS_VS_TRIS}: When a coarse triangulated mesh is optimized around a sharply peaked solution, the result can have a jagged shape where concavity varies greatly between neighboring triangles. After subdividing the triangles to achieve better resolution, the triangles within the concave region would have to be ``flipped'' to make the surface locally convex again. This problem is altogether avoided by using quads instead of triangles.

\subsection{Numerical optimization}
\label{ss_numerical_optimization}

We require a numerical method for finding a $D$-dimensional vector $\vec{x}=(x_1,\dots,x_D)$ extremizing a function $f(\vec{x})$, i.e.\ finding a point $\tilde{\vec{x}}$ with
\begin{equation}
\label{EQ_EXTREMUM}
\left. \frac{\partial f(\vec{x})}{\partial x_i} \right|_{\vec{x}=\tilde{\vec{x}}}= 0 \quad\quad \forall \; i \in (1,\dots,D) \text{ .}
\end{equation}
A textbook approach to this problem is Newton's method\footnote{
The term ``Newton's method'' is sometimes used exclusively for the original zero-point search method from which the extremum search method directly follows.} (see e.g.\ \cite{BurdenFairesCh10_2,UeberhuberCh14_4_1,PolakCh1_4}). This iterative method  starts from an initial point $\vec{x}^{(0)}$, onto which a recursion step is applied. This formula is given by
\begin{equation}
\label{EQ_NEWTONSTEP}
\vec{x}^{(n+1)} = \vec{x}^{(n)} - H[f(\vec{x}^{(n)})]^{-1} \vec{\nabla} f(\vec{x}^{(n)}) \text{ ,}
\end{equation}
with the Hessian matrix $H$ defined by
\begin{equation}
H[f(\vec{x})]_{i j} = \frac{\partial^2 f(\vec{x})}{\partial x_i \partial x_j} \text{ .}
\end{equation}
The formula \eqref{EQ_NEWTONSTEP} is computationally demanding for large $D$, as it requires an $O(D^2)$ computation of $H$ as well as an $O(D^3)$ matrix inversion. For our 4-dimensional problem with $N_V$ dynamic vertices describing the discretized geometry, we would set $D=4N_V$ for a naive implementation of Newton's method. Clearly, such an approach is not computationally efficient, and a variety of approximate quasi-Newton methods are used. For example, in \cite{Fonda:2014cca} the popular conjugate gradient method is used for similar problems on fixed time-slices.

Discretized surfaces in 4-dimensional spacetime live in a constrained region of the $4N_V$-dimensional parameter space we wish to probe numerically, as we have to avoid locally time-like solutions. On any quasi-Newton method that modifies several of the $4N_V$ degrees of freedom at once, this imposes a set of complicated dynamical constraints. To avoid this, we use Newton's method in its exact form but apply it locally, making the constraints  easier to implement.

The function to be extremized is the total area $A=\sum_k A_{\square_k}$ of all quads $\square_k$. While the total area depends on all $N_V$ dynamic vertices, varying the coordinates of one vertex $v^\mu_i$ alone only changes the area of quads that contain the vertex. In order to make the geometry locally extremal, we only need to apply Newton's method to one vertex at a time. In other words, instead of a set of global conditions for extremal total area we use an equivalent set of local conditions
\begin{equation}
\label{EQ_EXTREMUM_2}
g_i^\mu = \sum_{\square_k \ni v_i} \frac{\partial A_{\square_k}(\dots,v_i,\dots)}{\partial v_i^\mu} = 0 \quad\quad \forall \; i \in (1,\dots,N_V),\; \mu \in (t,x,y,z) \text{ ,}
\end{equation}
where the sum runs over all quads  $\square_k$ that contain the vertex $v_i$. The corresponding recursion step for Newton's method turns from \eqref{EQ_NEWTONSTEP} into
\begin{equation}
\label{EQ_NEWTONSTEP2}
\left(v_i^{(n+1)}\right)^\mu = \left(v_i^{(n)}\right)^\mu - \sum_{\square_k \ni v_i} \left[ \left( H[v_i^{(n)}]^{-1} \right)^{\mu \nu} \; \frac{\partial A_{\square_k}(v_i^{(n)})}{\partial (v_i^{(n)})^\nu} \right] \text{ ,}
\end{equation}
which is applied to each vertex 4-vector $v^\mu_i$ in the mesh. The Hessian is now only a $4\times4$ matrix which can be inverted without much computational cost. Upon introducing normal vectors, the Hessian is further simplified to a $2\times2$ matrix. If the recursion step leads to a locally time-like solution, the step size $\Vert v_i^{(n+1)} - v_i^{(n)} \Vert$ is reduced until the solution becomes space-like again.

The recursion step \ref{EQ_NEWTONSTEP2} is repeated until the local gradient norm $\Vert g_i \Vert$ is smaller than a threshold value.\footnote{
For an inhomogeneous discretization, it is preferable to apply the threshold to the gradient density $\Vert g_i \Vert / A_{v_i}$ instead, with $A_{v_i}=\sum_{\square_k \ni v_i} A_{\square_k}$ being the area of all quads containing the vertex $v_i$.} By repeatedly optimizing all vertices, the gradient norm vanishes across the entire mesh, with $A$ converging to the extremal area.

This approach of successively optimizing each vertex to extremize its local environment is computationally efficient only if applied to a geometry that is already close to the exact solution. Otherwise, the number of local updates needed to fulfill condition \eqref{EQ_EXTREMUM_2} for all vertices $v_i$ will become strongly nonlinear in $N_V$. Recursively subdividing and optimizing avoids such problems: The first few iterations establish the rough shape of the  solution by varying only a few dynamic vertices. Every subsequent iteration gradually refines the geometry, successively reducing the computational optimization cost per vertex.

Generally, Newton's method could be applied to the $N_V$ vertices in any order. However, we find it is more efficient to sort the vertices according to their local gradient norm $\Vert g_i \Vert$ and always optimize the vertex with the largest gradient. After the vertex is optimized, the vertices in the local neighborhood have their entries in the gradient list updated, and the process is repeated. This gradient update only has a computational cost of $O(\log N_V)$, but allows the algorithm to focus on the part of the mesh that is least extremal, speeding up convergence.

Our formulation of Newton's method is not fully covariant: The gradient vector, i.e.\ the covariant derivative of the scalar area functional, is equivalent to an ordinary partial derivative, so the fixed points of the method are independent of the coordinate system chosen. The Hessian, however, is not. While fully covariant descriptions of Newton's method exist \cite{Dedieu2003}, the computational cost of computing Christoffel symbols at every step does not make this approach favorable.

\subsection{Normals}

While there are now four dynamic parameters per vertex to optimize, we would like to constrain the optimization to directions orthogonal to the local surface. This is because moving the vertex along a tangent direction changes the resolution of the discretization, which can lead to clustering of vertices in one region of the mesh. To keep the discretization points at a distance, we
restrict Newton's method to directions along normal vectors.

\begin{figure}[t]
\centering
\includegraphics[width=0.5\textwidth]{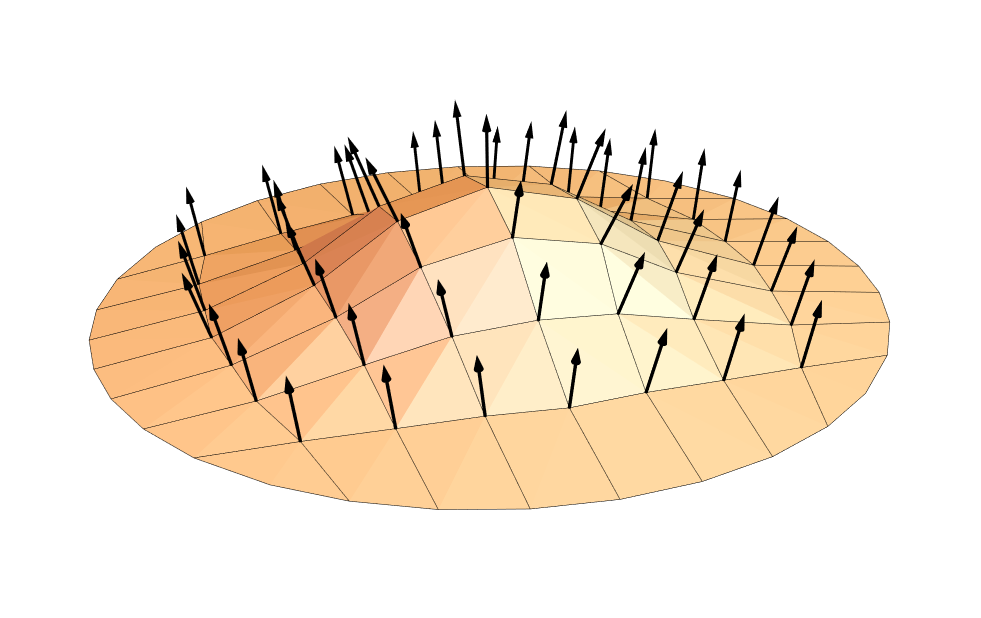}
\caption{A quadrilateralized surface with 3-dimensional normal vectors at each vertex. Boundary vertices do not have an associated normal vector.}
\label{FIG_QUAD_NORMALS}
\end{figure}

An example for 3-dimensional vertex normals is shown in figure \ref{FIG_QUAD_NORMALS}. Each dynamic (i.e.\ non-boundary) vertex lies along the corners of four quads, and each corner has an orthogonal direction given by the vector product of the two edges involved. Averaging those over all corners, an effective vertex normal is produced.

In curved 4-dimensional spacetime, we need to consider differential edges in the local neighborhood of each vertex. Furthermore, a 2-dimensional surface in four dimensions has a 2-dimensional space of 4-vectors orthogonal to each point on the surface, so normal vectors cannot be uniquely defined. We are therefore free to choose a pair of 4-dimensional (differential) normals $\text{d}n_\mu$ and $\text{d}m_\mu$ in the following manner:
\begin{equation}
\label{EQ_NORMALS1}
\text{d}n_\mu = \epsilon_{\mu \nu \rho \sigma}\; \text{d}v^\nu \text{d}w^\rho \text{d}t^\sigma \quad \text{ , }\quad \text{d}m_\mu = \epsilon_{\mu \nu \rho \sigma}\; \text{d}v^\nu \text{d}w^\rho \text{d}n^\sigma \text{ .}
\end{equation}
Here, $\text{d}v^\mu$ and $\text{d}w^\mu$ are two differential vectors spanning a corner of a quad, and $\epsilon_{\mu \nu \rho \sigma}$ is the 4-dimensional Levi-Civita symbol. The constant vector $\text{d}t^\mu$ is an arbitrary covariant time-like vector. Using differential forms, we can also write \eqref{EQ_NORMALS1} more generally as
\begin{equation}
\label{EQ_NORMALS2}
\text{d}n = \star \left( \text{d}v \land \text{d}w \land \text{d}t \right) \quad \text{ , }\quad \text{d}m = \star \left( \text{d}v \land \text{d}w \land \text{d}n \right) \text{ ,}
\end{equation}
where $\text{d}t$ is now a differential 1-form. In practice, we can simply set
$\text{d}t=\delta_{0 \mu} \text{d}x^\mu$.

In this construction, $\text{d}n_\mu$ is a space-like vector and $\text{d}m_\mu$ a time-like one. Thus, an extremal surface will be minimal along the former and maximal along the latter direction. For a surface at constant time, i.e.\ with $\text{d}v^0=\text{d}w^0=0$, the first normal $\text{d}n_\mu$ corresponds to the usual 3-dimensional normal vector with a vanishing time component while the second normal $\text{d}m_\mu$ only points in the time direction.

For computational purposes, normal vectors are simply vectors of unit length, as we are only interested in the direction of the normals. Due to this rescaling, we have omitted a factor $\sqrt{|\det g|}$ in \eqref{EQ_NORMALS1} that would appear in a covariant definition of the Levi-Civita symbol.

\subsection{Time-slice constraints}
\label{ss_timeslice_constraints}

For strongly time-dependent solutions, it may not be possible during the first few iterations to find a low-resolution mesh that follows the general shape of the smooth solution. This can lead to unstable conditions after subdividing the quad mesh, such as time-like quads. This problem can be avoided using a series of preconditioning steps, during which we restrict the mesh to a time-slice compatible with the boundary conditions. When the mesh has become sufficiently smooth, the time-slice constraint is dropped.

A suitable choice of the metric $g_{\mu \nu}$ also increases the stability of the algorithm. Assume the initial coordinates are $(t,x,y,z)$. If the mesh boundary lies on some space-like surface given by $t = f(x,y,z)$, we can simply introduce a modified time coordinate $t \to t^\prime = t - f(x,y,z)$. In the new $(t^\prime,x,y,z)$ coordinates, we can now construct an initial mesh at $t^\prime=0$. Under the new metric, tangent vectors along any point on this mesh will lie on the space-like surface, so we can perform a time-slice constrained optimization without encountering time-like solutions. After these preconditioning steps, we allow the algorithm to optimize along the $t^\prime$ coordinate, as well.

\subsection{Accuracy and error estimation}

The accuracy of the area of the extremized surface mainly depends on two parameters, the gradient convergence threshold and the number of iterations. The threshold value serves to terminate both the local steps of Newton's method as well as the entire iteration, where in the latter case it is applied to the maximum gradient along the whole mesh. The absolute value of this threshold should be decreased with each iteration step, so that smoothness on increasingly smaller scales is achieved. For our purposes, we used a threshold gradient density of $10^{-2}$.

\begin{figure}[t]
\centering
\includegraphics[width=0.5\textwidth]{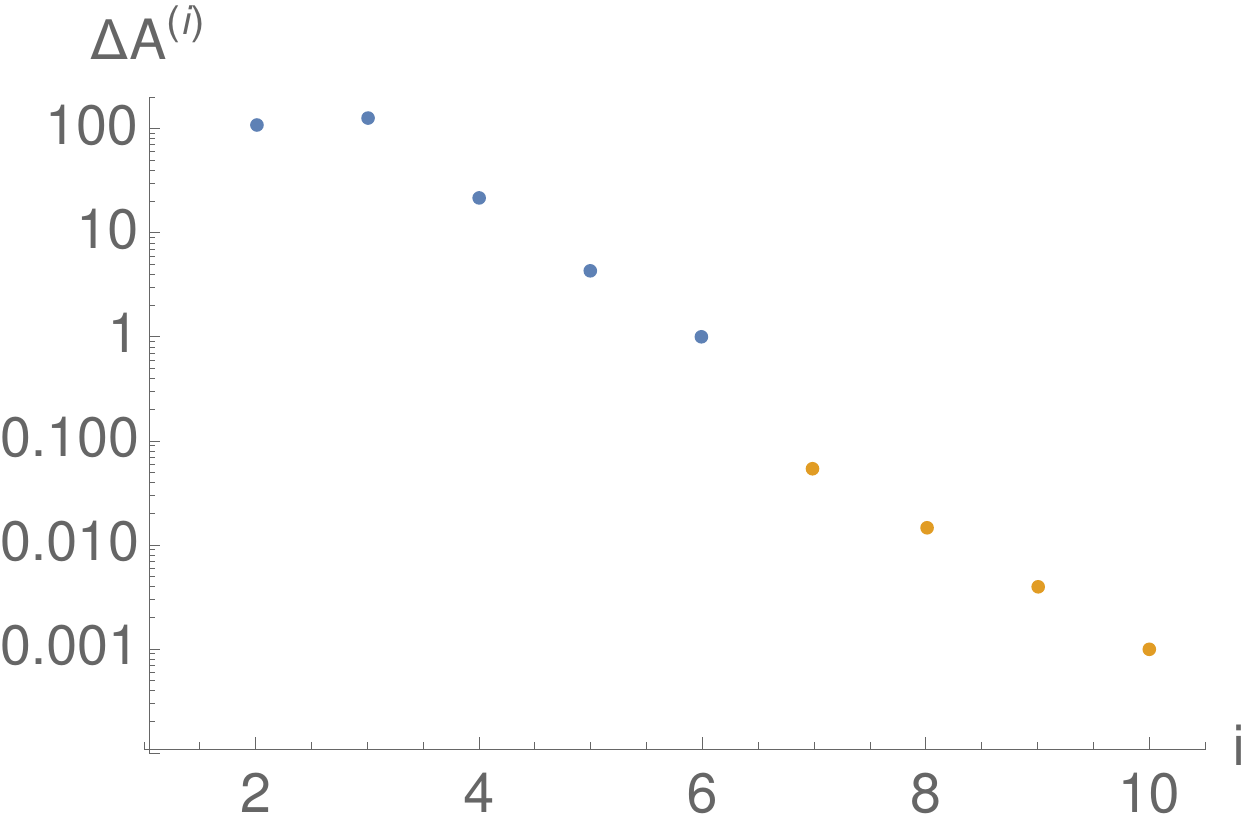}
\caption{Typical convergence of the area of an extremal surface. $\Delta A^{(i)} = | A^{(i)} - A^{(i-1)}|$ is the difference of the area between subsequent iterations $i-1$ and $i$. The blue dots correspond to results under a time-slice constraint, which is dropped at $i=6$. The data was computed for the AdS$_4$ local quench setup considered in the main text, at time $t=e^{3.5}$ and mass parameter $M=1.28$ for a boundary with proper distance $l=3.5$ from the coordinate horizon. Units in $R=\alpha=1$.}
\label{FIG_AREA_CONV}
\end{figure}

Given a sufficiently small gradient threshold, the area of the discretized surface converges exponentially with the number of iterations, as each iteration increases the area resolution by a factor of four. Denote as $\Delta  A^{(i)} = | A^{(i)} - A^{(i-1)}|$ the difference between the extremal surface area at iteration $i$ and $i-1$. If $A^{(i)}$ converges exponentially with $i$, so does $\Delta A^{(i)}$. An example for such a convergence is shown in figure \ref{FIG_AREA_CONV}. Note that the convergence is more erratic during earlier iterations, as the extremal surface still changes shape considerably. After a few iterations, however, the convergence quickly becomes exponential. Taking the differences $\Delta A^{(N_i)}$ and $\Delta A^{(N_i-1)}$  between the last three iterations, and assuming that $\Delta A^{(i)}$ continues to converge exponentially, we can estimate the absolute error $\delta A^{(N_i)} = |A^{(\infty)} - A^{(N_i)}|$ after $N_i$ iterations as
\begin{equation}
\label{EQ_AREA_ERROR}
\delta A^{(N_i)} \simeq \sum_{n=1}^\infty \Delta A^{(N_i)} \left( \frac{\Delta A^{(N_i-1)}}{\Delta A^{(N_i)}} \right)^{-n} = \frac{\left(\Delta A^{(N_i)}\right)^2}{\Delta A^{(N_i-1)} - \Delta A^{(N_i)}} \text{ ,}
\end{equation}
where we sum up the projected steps of all further iterations. For example, the data in figure \ref{FIG_AREA_CONV} leads to an estimated error $\delta A^{(10)} \approx 3 \cdot 10^{-4}$. If $A^{(i)}$ does not converge monotonically (but $\Delta A^{(i)}$ still decreases exponentially), \eqref{EQ_AREA_ERROR} gives an upper bound to the absolute error.

When computing the differences of surfaces in the limit of infinitely large boundaries, we also need to consider the results for different effective radii $l$. As long as $l$ is chosen to correspond to some proper length under the given metric, these results typically converge exponentially with $l$ as well, so we can compute errors similar to \eqref{EQ_AREA_ERROR}.

\subsection{Performance}

As the computational cost of one local update is constant, the performance of the algorithm scales linearly with the number of local updates required to reach the gradient convergence threshold. The iterative approach of optimizing the mesh at gradually higher resolutions implies that optimizing one vertex only affects the gradient of vertices within an effective local region $A_\text{loc} \propto \left( \frac{1}{4} \right)^{k}$ of the entire mesh at the $k$th iteration. Because the number of vertices within $A_\text{loc}$ remains independent of $k$, the necessary number of local updates per vertex is constant as well. Thus, the performance of the algorithm should scale linearly in the number of vertices to be optimized, i.e.\ with $O(4^{N_i})$, $N_i$ being the number of total iterations. In reality, the actual performance drop is slightly higher, as the memory requirements also increase exponentially, reducing access times in any practical implementation. Note that processes such as normal construction and mesh subdivision have no noticeable impact on total runtime, as they are only executed once per iteration.

For producing the data presented in this paper, we used up to $N_i=10$ iterations, which corresponds to $\sim 10^6$ discretization points and as many quads. On a typical office CPU, one such calculation takes several days to complete and requires up to $\sim 1\mathrm{GB}$ of RAM.

\end{document}